\documentclass[10pt]{iopart}
\usepackage{graphicx}
\usepackage{tabls}
\eqnobysec

\newcommand{\beq}{\begin{equation}}
\newcommand{\beqa}{\begin{eqnarray}}
\newcommand{\eeq}{\end{equation}}
\newcommand{\eeqa}{\end{eqnarray}}

\newcommand{\bra}[1]{\langle#1\vert}
\newcommand{\ket}[1]{\vert#1\rangle}
\newcommand{\braket}[2]{\langle#1\vert#2\rangle}
\newcommand{\braopket}[3]{\langle#1\vert#2\vert#3\rangle}

\renewcommand{\a}{\alpha}
\newcommand{\abs}[1]{\vert#1\vert}
\renewcommand{\bar}[1]{{\overline{#1}}}
\newcommand\dbl[2]{{\matrix{\hfill\mbox{#1}\hfill\cr\hfill\mbox{#2}\hfill}}}
\newcommand{\dd}{{\rm d}}
\renewcommand{\dj}{d^{(j)}}
\newcommand{\ds}{d^{(S)}}
\newcommand{\edge}{{{\rm edge}}}
\newcommand{\ecal}{{\cal E}}
\newcommand{\eps}{\varepsilon}
\newcommand{\esp}{{{\hskip 17pt}}}
\newcommand\frad[2]{{\displaystyle{#1\over #2}}}
\newcommand\frat[2]{{\textstyle{#1\over #2}}}
\newcommand{\half}{\frat{1}{2}}
\newcommand{\ical}{{\cal I}}
\newcommand{\ii}{{\rm i}}
\renewcommand{\max}{{{\rm max}}}
\renewcommand{\min}{{{\rm min}}}
\newcommand{\mean}[1]{\langle#1\rangle}
\newcommand{\phij}{\varphi^{(j)}}
\renewcommand{\th}{\theta}
\newcommand{\var}{\mathop{\rm var}\,}
\newcommand{\vv}{{\vphantom{X}}}

\newcommand{\Ai}{{\mathop{\rm Ai}}}
\renewcommand{\H}{{\cal H}}
\newcommand{\K}{{\bf K}}
\renewcommand{\Re}{\mathop{\rm Re}\,}
\renewcommand{\O}{{\rm O}}
\newcommand{\U}{{\rm U}}

\begin{document}

\title{Equilibration properties of small quantum systems: further examples}

\author{J M Luck}

\address{Institut de Physique Th\'eorique, Universit\'e Paris-Saclay, CEA and CNRS,
91191~Gif-sur-Yvette, France}

\begin{abstract}
It has been proposed to investigate the equilibration properties
of a small isolated quantum system by means of the matrix
of asymptotic transition probabilities in some preferential basis.
The trace $T$ of this matrix measures the degree of equilibration of the system
prepared in a typical state of the preferential basis.
This quantity may vary between unity (ideal equilibration)
and the dimension $N$ of the Hilbert space (no equilibration at all).
Here we analyze several examples of simple systems
where the behavior of $T$ can be investigated by analytical means.
We first study the statistics of $T$
when the Hamiltonian governing the dynamics is random and drawn
from a distribution invariant under the group U$(N)$ or O$(N)$.
We then investigate a quantum spin $S$ in a tilted magnetic field
making an arbitrary angle with the preferred quantization axis,
as well as a tight-binding particle on a finite electrified chain.
The last two cases provide examples of
the interesting situation where varying a system parameter
-- such as the tilt angle or the electric field --
through some scaling regime
induces a continuous crossover from good to bad equilibration properties.
\end{abstract}

\ead{\mailto{jean-marc.luck@cea.fr}}

\maketitle

\section{Introduction}

The last decade has witnessed an immense activity
around the themes of therma\-lization and equilibration in isolated quantum systems
(see~\cite{cr,pss,efg,nh,ge,akp,santosrev} for comprehensive reviews).
This renewed interest in an old classic subject of Quantum Mechanics
was mostly triggered by progress on the experimental side in cold-atom physics.
The physical mechanisms underling equilibration and thermalization
are far from obvious.
On the one hand, an isolated quantum system evolves by some unitary dynamics,
and therefore keeps the memory of its initial state.
On the other hand, if the system is large enough,
concepts from Statistical Mechanics can be expected to apply at least approximately.
Most recent works have dealt with large many-body quantum systems,
the key issues including the description
of a small subsystem by a thermodynamical ensemble
(either microcanonical or canonical),
the thermalization properties of single highly-excited states,
and the role of conservation laws and of integrability.

An alternative and complementary viewpoint,
where the main focus is on the unitary evolution
of a small quantum system considered as a whole,
has been emphasized in~\cite{I}.
There, the key quantity is the matrix $Q$ of asymptotic transition probabilities
in some preferential basis.
The trace $T$ of this matrix is also the sum of the inverse participation ratios
of all eigenstates of the Hamiltonian in the same basis.
This quantity has been put forward as a measure of the degree
of equilibration of the full system if launched from a typical basis state;
the larger $T$, the poorer the equilibration.

The above line of thought has been illustrated
by means of a detailed study of a single tight-binding particle
on a finite segment of $N$ sites in a random potential~\cite{I}.
In the absence of disorder,
the quantity $T$ exhibits a finite value $T=T^{(0)}\approx 3/2$,
testifying good equilibration.
In the localized regime, $T\approx N\bar{Q}$ grows linearly with the sample size,
with a finite mean asymptotic return probability $\bar{Q}$,
testifying poor equilibration.
The outcome of most physical interest concerns the regime of a weak disorder,
where the disorder strength $w$ is much smaller than the bandwidth.
In this regime, $T$ obeys a universal scaling law of the form
\beq
T\approx T^{(0)}+\frac{1}{N}\,\Phi(x),\qquad x=Nw^{2/3}.
\label{tfss}
\eeq
This finite-size scaling law interpolates between both above regimes.
In the ballistic regime, where $x$ is small,
the first correction due to disorder scales as $\Phi(x)\approx 13x^3/540$.
When $x$ is large, the quadratic growth $\Phi(x)\approx Ax^2$,
with $A\approx0.21$, describes the crossover to the localized regime.
The mean return probability accordingly scales as $\bar{Q}\approx Aw^{4/3}$
at weak disorder.
The crossover exponent 2/3 entering
the variable~$x$ is dictated by the well-known anomalous scaling
of the localization length in the vicinity of band edges.
Typical equilibration properties of a single particle in a weak random potential,
as testified by the scaling behavior of $T$,
are therefore governed by the few most localized band-edge states.
This is a not an artifact of considering a global quantity such as $T$,
albeit a genuine physical effect:
a typical initial state will generically have a non-zero overlap
with the anomalously localized band-edge states,
and the very slow relaxation of the latter overlap will cause
the relatively poor equilibration properties of such a generic initial state.

The behavior of $T$ in a clean quantum many-body system has then been considered
in~\cite{mpl}, where the statistics of inverse participation ratios in the XXZ spin chain
with anisotropy parameter $\Delta$ has been investigated numerically.
There, the quantity $T$ was found to grow linearly with the system size ($T\sim L$)
in the gapless phase of the model ($\abs{\Delta}<1$),
and exponentially ($T\sim\exp(aL)$) in the gapped phase ($\abs{\Delta}>1$).
In the particular case of the XX spin chain ($\Delta=0$),
which can be mapped onto a free-fermion problem,
the inverse participation ratios of some individual eigenstates
have been investigated in~\cite{xx}.
The amount of technical effort needed to derive the partial results presented there
attests the high level of difficulty of the analysis of seemingly simple
quantities such as $T$, even in the simplest of integrable models.

In this paper we analyse in detail several other examples of simple quantum systems
for which the behavior of $T$ can be investigated by analytical means.
Section~\ref{reminder} provides a reminder of the transition matrix formalism
proposed in~\cite{I}, emphasizing the role and the meaning of $T$.
In section~\ref{rmt} we investigate the statistics of $T$
corresponding to a random Hamiltonian
whose distribution is invariant under the group U$(N)$ or O$(N)$,
thus keeping in line with the random matrix theory approach to quantum chaos.
Section~\ref{spin} deals with a single quantum spin $S$
submitted to a tilted magnetic field
making an angle $\th$ with respect to the quantization axis defining the preferred basis.
Finally, the case of a tight-binding particle on a finite electrified chain
is considered in section~\ref{stark}.
This situation is somehow a deterministic analogue of the case considered in~\cite{I},
where the ladder of Wannier-Stark states replaces Anderson localized states.
It is however richer,
as it features two successive scaling regimes in the small-field region.
Our findings are summarized and discussed in section~\ref{disc}.

\section{A reminder on the transition matrix formalism}
\label{reminder}

This section provides a reminder of the transition matrix formalism
which has been proposed in~\cite{I}
to describe equilibration in small isolated quantum systems.
Consider a system whose Hilbert space has dimension~$N$.
Let $\{\ket{a}\}$ ($a=1,\dots,N$) be a preferential basis, chosen once for all.
In that basis the Hamiltonian $\H$ is some $N\times N$ matrix.
We assume that its eigenvalues $E_n$ are non-degenerate.
This condition will be satisfied in all systems considered in this work.
The situation of degenerate energy eigenvalues will be considered briefly
at the end of this section.

If the system is launched from one of the basis states, say $\ket{a}$,
the probability for it to be in state $\ket{b}$ at a subsequent time $t$ reads
\beq
P_{ab}(t)
=\sum_{m,n}\e^{\ii(E_n-E_m)t}\braket{b}{m}\braket{m}{a}\braket{a}{n}\braket{n}{b},
\label{pab}
\eeq
where $\{\ket{n}\}$ ($n=1,\dots,N$) is a basis of normalized eigenstates,
such that $\H\ket{n}=E_n\ket{n}$.
The stationary state reached by the system
is characterized by the time-averaged transition probabilities
\beq
Q_{ab}=\lim_{t\to\infty}\frac{1}{t}\int_0^tP_{ab}(t')\,\dd t'.
\eeq
In the absence of spectral degeneracies, we have
\beq
Q_{ab}=\sum_n\abs{\braket{a}{n}}^2\;\abs{\braket{b}{n}}^2.
\label{qab}
\eeq

The transition matrix $Q$ is the central object of this formalism.
Let us begin with a few general properties.
The matrix $Q$ only depends on the eigenstates $\ket{n}$ of the Hamiltonian $\H$,
and not on its eigenvalues $E_n$.
Therefore, any two Hamiltonians with the same eigenstate basis
share the same matrix $Q$, irrespective of their spectra.
In full generality, the matrix $Q$ reads
\beq
Q=RR^T,
\label{rrt}
\eeq
where the matrix $R$ is given by
\beq
R_{an}=\abs{\braket{a}{n}}^2,
\label{ran}
\eeq
and $R^T$ is its transpose.
The matrix $Q$ is therefore real symmetric and positive definite.
The matrix $R$ is real, albeit not symmetric in general;
it may therefore have complex spectrum.
Finally, both matrices $Q$ and $R$ are doubly stochastic:
their row and column sums equal unity:
\beq
\sum_aR_{an}=\sum_nR_{an}=\sum_aQ_{ab}=\sum_bQ_{ab}=1.
\eeq
More generally, one can associate real matrices $Q$ and $R$ obeying the above properties
with any pair of orthonormal bases $\{\ket{a}\}$ and $\{\ket{n}\}$,
related to each other by an arbitrary unitary transformation $U$ such that
\beq
U_{an}=\braket{a}{n}.
\label{udef}
\eeq

Coming back to equilibration properties, the diagonal element
\beq
Q_{aa}=\sum_n\abs{\braket{a}{n}}^4,
\label{qaa1}
\eeq
i.e., the stationary return probability to the basis state $\ket{a}$,
can be recast as
\beq
Q_{aa}=\tr\omega_a^2,
\label{qaa2}
\eeq
where
\beq
\omega_a=\sum_n\abs{\braket{a}{n}}^2\;\ket{n}\bra{n}
\label{oares}
\eeq
is the stationary density matrix issued from state~$\ket{a}$.
The right-hand side of~(\ref{qaa2}) is called the purity of $\omega_a$.
Its reciprocal, i.e.,
\beq
d_a=\frac{1}{Q_{aa}},
\eeq
provides a measure of the dimension of the subspace of Hilbert space
where the system equilibrates if launched from state~$\ket{a}$~\cite{prf,lps,lp,gpe,as,iu}.

The equilibration properties of the system launched from a typical basis state
are encoded in the trace of the matrix $Q$:
\beq
T=\sum_{a,n}\abs{\braket{a}{n}}^4=\sum_aQ_{aa}=\sum_nI_n,
\label{tdef}
\eeq
where
\beq
I_n=\sum_a\abs{\braket{a}{n}}^4
\label{iprdef}
\eeq
is the inverse participation ratio (IPR)~\cite{bdh,bd,vis,th}
of the eigenstate $\ket{n}$ of $\H$ in the preferential basis.
The return probabilities $Q_{aa}$ and the IPR $I_n$ therefore somehow
appear as dual to each other.

The ratio
\beq
D=\frac{N}{T}
\label{ddef}
\eeq
is such that
\beq
\frac{1}{D}=\frac{1}{N}\sum_a\frac{1}{d_a}.
\eeq
Hence $D$ provides a measure of the dimension of the subspace
which hosts the equilibration dynamics for a typical initial state.
The trace $T$ and the effective dimension $D$ are always
between the following extremal values.

\subsubsection*{Ideal equilibration.}

If the eigenstates of $\H$ are uniformly spread over the preferential basis, we have
\beq
R_{an}=\frac{1}{N}
\eeq
for all $a$ and $n$.
The bases $\{\ket{a}\}$ and $\{\ket{n}\}$ are said
to be mutually unbiased~\cite{vou,kibler,bz}.
Hence
\beq
Q_{ab}=\frac{1}{N}
\eeq
for all $a$ and $b$,
and
\beq
T_\min=1,\qquad D_\max=N.
\label{tmin}
\eeq
This ideal situation corresponds to a perfectly equilibrated stationary state,
with no memory of the initial state at all.

\subsubsection*{No equilibration.}

This situation is the exact opposite of the previous one.
If $\H$ is diagonal in the preferential basis,
its eigenstates $\ket{n}$ can be ordered so as to have
\beq
R_{an}=\delta_{an}
\eeq
for all $a$ and $n$, and
\beq
Q_{ab}=\delta_{ab}
\eeq
for all $a$ and $b$,
$\delta$ being the Kronecker symbol,
and so
\beq
T_\max=N,\qquad D_\min=1.
\label{tmax}
\eeq
In such a situation, the system keeps a full memory of its initial state,
and so there is no equilibration at all.

\medskip

Finally, let us mention that the above transition matrix formalism
can be readily extended to situations
where some of the energy eigenvalues are degenerate~\cite{I}.
In such a situation, the expression~(\ref{tdef}) of the quantity $T$ becomes
\beq
T=\sum_{a,n}\left(\sum_{i=1}^{\mu_n}\abs{\braket{a}{n,i}}^2\right)^2,
\label{dtdef}
\eeq
where the $\ket{n,i}$ form an orthonormal basis such that
$\H\ket{n,i}=E_n\ket{n,i}$, for $i=1,\dots,\mu_n$,
where $\mu_n$ is the multiplicity (degeneracy)
of the $n$th energy eigenvalue~$E_n$.

\section{Random matrix theory approach}
\label{rmt}

In this section we investigate the statistics of $T$
obtained by choosing the Hamiltonian matrix $\H$ at random,
according to some probability distribution over $N\times N$ Hermitian matrices.
We assume that the latter distribution is invariant
under the action of a suitable chosen symmetry group, here U$(N)$ or O$(N)$
(see below).
This assumption keeps in line with the random matrix theory approach to quantum chaos,
which has been used extensively to investigate many properties
of complex quantum systems~\cite{bee,gmw,mehta,abd},
including the energy spectra of closed systems
and the scattering and transport properties of open systems.
This line of thought implies in particular that the distribution of the matrix $U$,
introduced in~(\ref{udef}),
encoding the eigenstates of $\H$ in the preferred basis,
is given by the flat Haar measure on the appropriate symmetry group.
Moreover, it will be sufficient for our purpose to consider the first two cumulants of $T$,
i.e., the mean value $\mean{T}$ and the variance $\var T=\mean{T^2}-\mean{T}^2$.

\subsubsection*{The unitary case.}

We consider first the situation
where the Hamiltonian $\H$ is an $N\times N$ complex Hermitian matrix.
The relevant symmetry group is then the unitary group U$(N)$.
We have
\beqa
\esp\mean{T}_\U&=&N^2A_\U,
\\
(\var T)_\U&=&N^2\left(B_\U+2(N-1)C_\U+(N-1)^2D_\U-N^2A_\U^2\right),
\eeqa
with
\beq
\matrix{
A_\U=\mean{\abs{U_{11}}^4}_\U,\hfill&
B_\U=\mean{\abs{U_{11}}^8}_\U,\hfill\cr
C_\U=\mean{\abs{U_{11}}^4\abs{U_{12}}^4}^\vv_\U,\hfill\qquad&
D_\U=\mean{\abs{U_{11}}^4\abs{U_{22}}^4}_\U,\hfill\cr
}
\label{abcdu}
\eeq
where $\mean{\dots}_\U$ denotes an average
with respect to the Haar measure on~U$(N)$.

Moments of this kind have attracted much attention in the past,
and several approaches have been proposed
to evaluate them~\cite{u,wei,iz,sam,dz,bc,bcs,al,db,gl}.
Table~\ref{list} gives the values of the moments
$A_\U$, $B_\U$, $C_\U$ and $D_\U$
for the unitary group U$(N)$ (see~(\ref{abcdu}))
and $A_\O$, $B_\O$, $C_\O$ and $D_\O$
for the orthogonal group O$(N)$ (see~(\ref{abcdo})).

\begin{table}[!ht]
\begin{center}
\begin{tabular}{|c|c|}
\hline
U$(N)$ & O$(N)$\\
\hline
$A_\U=\frac{2^\vv}{N(N+1)}\hfill$&
$A_\O=\frac{3}{N(N+2)}\hfill$\\
$B_\U=\frac{24^\vv}{N(N+1)(N+2)(N+3)}\hfill$&
$B_\O=\frac{105}{N(N+2)(N+4)(N+6)}\hfill$\\
$C_\U=\frac{4^\vv}{N(N+1)(N+2)(N+3)}\hfill$&
$C_\O=\frac{9}{N(N+2)(N+4)(N+6)}\hfill$\\
$D_\U=\frac{4^\vv}{N^2(N-1)(N+3)_\vv}\hfill$&
$D_\O=\frac{9(N+3)(N+5)}{N(N^2-1)(N+2)(N+4)(N+6)}\hfill$\\
\hline
\end{tabular}
\end{center}
\caption{\small Values of the integrals $A$, $B$, $C$ and $D$
for both symmetry classes.
For the unitary group U$(N)$ (see~(\ref{abcdu})),
the one-vector integrals $A_\U$, $B_\U$ and $C_\U$ are given in~\cite{u,gl}
while the two-vector integral $D_\U$, given in~\cite{al},
can be checked using the recursive approach of~\cite{gl}.
For the orthogonal group O$(N)$ (see~(\ref{abcdo})),
the one-vector integrals $A_\O$, $B_\O$ and $C_\O$ are given in~\cite{gl},
and the two-vector integral $D_\O$, given in~\cite{db},
can again be checked using~\cite{gl}.}
\label{list}
\end{table}

Using the results of table~\ref{list}, we obtain
\beqa
\esp\mean{T}_\U&=&\frac{2N}{N+1},
\label{tu}
\\
(\var T)_\U&=&\frac{4(N-1)}{(N+1)^2(N+3)}.
\label{vtu}
\eeqa

\subsubsection*{The orthogonal case.}

We now turn to the situation where the dynamics is invariant under time reversal,
so that the Hamiltonian $\H$ is a real symmetric matrix.
The relevant symmetry group is then the orthogonal group~O$(N)$.
We have
\beqa
\esp\mean{T}_\O&=&N^2A_\O,
\\
(\var T)_\O&=&N^2\left(B_\O+2(N-1)C_\O+(N-1)^2D_\O-N^2A_\O^2\right),
\eeqa
with
\beq
\matrix{
A_\O=\mean{\Omega_{11}^4}_\O,\hfill&
B_\O=\mean{\Omega_{11}^8}_\O,\hfill\cr
C_\O=\mean{\Omega_{11}^4\Omega_{12}^4}^\vv_\O,\hfill\qquad&
D_\O=\mean{\Omega_{11}^4\Omega_{22}^4}_\O,\hfill\cr
}
\label{abcdo}
\eeq
where $\mean{\dots}_\O$ now denotes an average
with respect to the Haar measure on~O$(N)$.
Using again the results of table~\ref{list}, we obtain
\beqa
\esp\mean{T}_\O&=&\frac{3N}{N+2},
\label{to}
\\
(\var T)_\O&=&\frac{24N(N-1)}{(N+1)(N+2)^2(N+6)}.
\label{vto}
\eeqa

\medskip

For a very large dimension $N$,
$T$ converges to the limits
\beq
T_\U=2,\qquad T_\O=3,
\label{tls}
\eeq
which only depend on the symmetry class of the Hamiltonian.
These finite values testify extended eigenstates
and good equilibration properties,
as could be expected for a generic Hamiltonian.
The limit values~(\ref{tls}) are deterministic,
in the sense that fluctuations of individual values of $T$ become negligible
for large $N$, as testified by the $1/N^2$ falloff of the variances
\beq
(\var T)_\U\approx\frac{4}{N^2},\qquad
(\var T)_\O\approx\frac{24}{N^2}.
\label{tvars}
\eeq

The limit values~(\ref{tls}) coincide with the outcome of the following naive approach.
In the unitary case,
consider $T$ as the sum of $N^2$ independent quantities of the form~$\abs{u}^4$,
modeling the entries $u$ as centered complex Gaussian variables
such that $\mean{\abs{u}^2}=1/N$.
The identity $\mean{\abs{u}^4}=2\mean{\abs{u}^2}^2$
yields $\mean{\abs{u}^4}=2/N^2$,
and so $T_\U=N^2\mean{\abs{u}^4}=2$.
Similarly, in the orthogonal case,
consider $T$ as the sum of~$N^2$ independent quantities of the form~$\omega^4$,
modeling the entries $\omega$ as real Gaussian variables
such that $\mean{\omega^2}=1/N$.
The identity $\mean{\omega^4}=3\mean{\omega^2}^2$
yields $\mean{\omega^4}=3/N^2$,
and so $T_\O=N^2\mean{\omega^4}=3$.
In the same vein, the $1/N^2$ decay of the variances~(\ref{tvars})
also conforms with the above naive approach,
viewing $T$ as the sum of $N^2$ independent random variables.

\begin{figure}[!ht]
\begin{center}
\includegraphics[angle=-90,width=.475\linewidth]{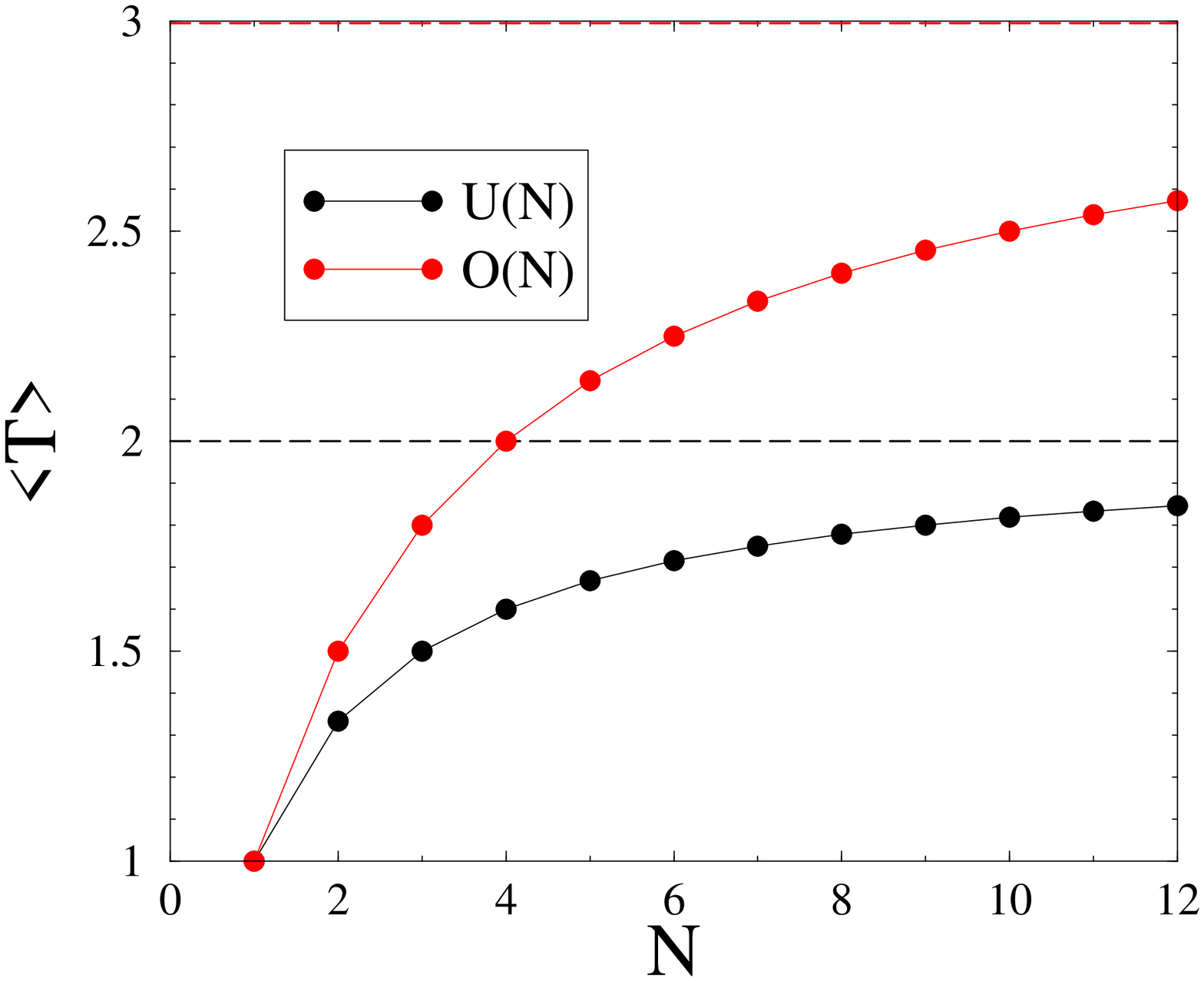}
\includegraphics[angle=-90,width=.475\linewidth]{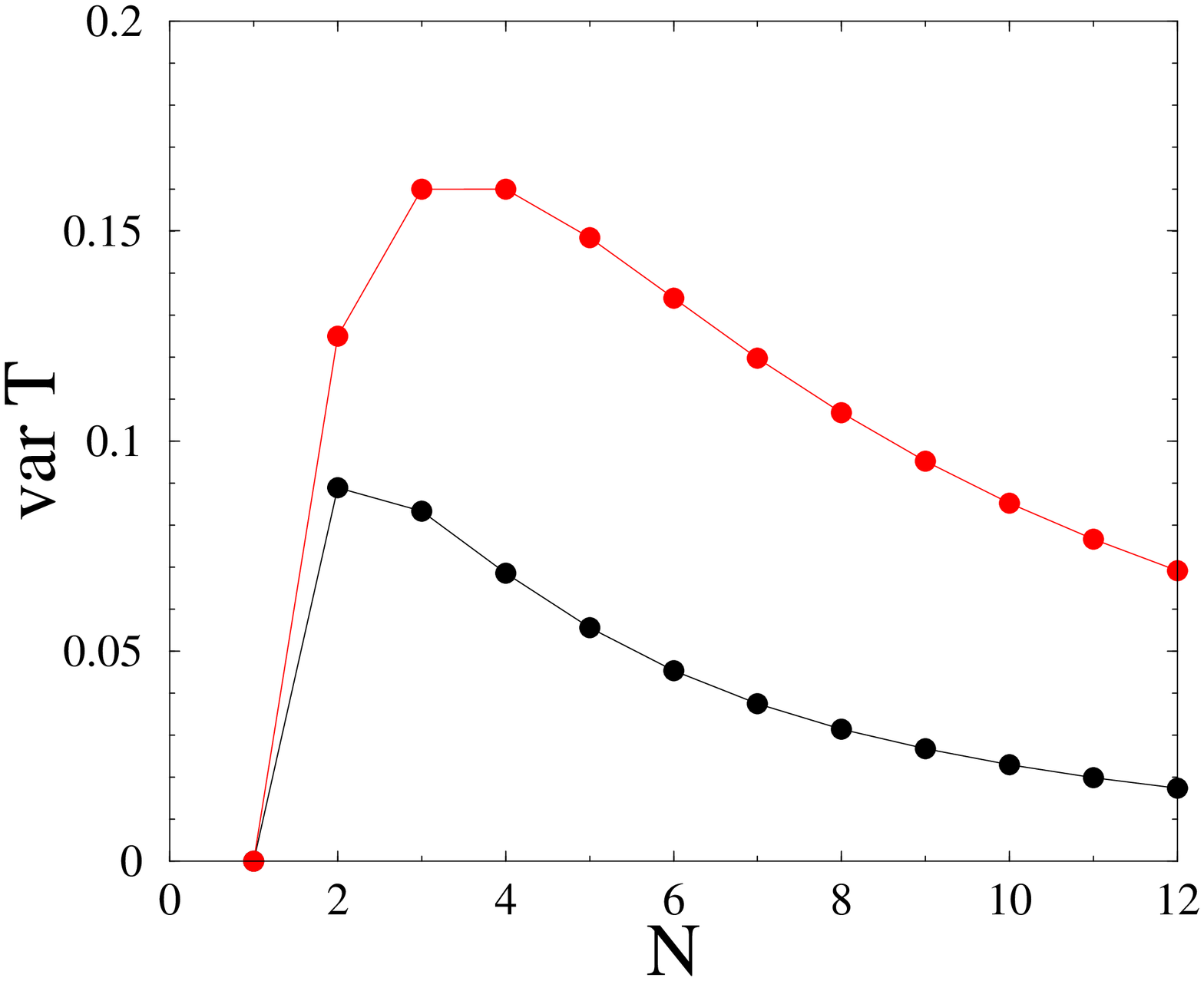}
\caption{\small
Mean values (left) and variances (right) of $T$,
for a random Hamiltonian with a distribution invariant under
the unitary group U$(N)$ (see~(\ref{tu}),~(\ref{vtu}))
or the orthogonal group O$(N)$ (see~(\ref{to}),~(\ref{vto})).
Dashed horizontal lines in left panel:
large-$N$ limits $T_\U=2$, $T_\O=3$ (see~(\ref{tls})).}
\label{trmt}
\end{center}
\end{figure}

For finite values of $N$ (see figure~\ref{trmt}),
the results~(\ref{tu}),~(\ref{vtu}) and~(\ref{to}),~(\ref{vto})
have a simple rational dependence on~$N$.
The mean values $\mean{T}$ increase monotonically from unity
and saturate to the limits~(\ref{tls}).
The variances vanish both for $N=1$ and $N\to\infty$,
and therefore present a maximum.
The maximum of $(\var T)_\U$, equal to 4/45, is reached for $N=2$,
while the maximum of $(\var T)_\O$, equal to 4/25, is reached for $N=3$ and 4.

\section{A spin in a tilted magnetic field}
\label{spin}

This section is devoted to the dynamics of a single quantum spin
in the representation of SU$(2)$ of dimension
\beq
N=2S+1.
\eeq
The spin $S$ is either an integer (if $N$ is odd)
or a half-integer (if $N$ is even).
We choose the $z$-axis as preferred quantization axis.
The preferred basis thus consists of the eigenstates $\ket{a}$ of $S_z$,
where $a$ takes the~$N$ values $a=-S,\dots,S$, with unit steps.
The spin is subjected to a tilted external magnetic field~${\bf h}$
making an angle $\th$ with the $z$-axis, with $0\le\th\le\pi$.
The corresponding Hamiltonian reads, in dimensionless units
\beq
\H=\cos\th\,S_z+\sin\th\,S_x.
\eeq
The spin $S$ and the tilt angle $\th$ are the two parameters of the model.

\subsection{The simplest case: $S=\half$}

Let us consider first the case of a spin $S=\half$, so that $N=2$.
The Hamiltonian
\beq
\H=\frac{1}{2}\pmatrix{\cos\th&\sin\th\cr\sin\th&-\cos\th}
\eeq
has eigenvalues $E_\pm=\pm\half$.
The matrix encoding the corresponding eigenstates is
\beq
U=\pmatrix{\cos\frac{\th}{2}&\sin\frac{\th}{2}\cr
-\sin\frac{\th}{2}&\cos\frac{\th}{2}},
\eeq
up to arbitrary phases.
The matrices $R$ and $Q$ (see~(\ref{qab}),~(\ref{ran})) therefore read
\beqa
R&=&\frac{1}{2}\pmatrix{1+\cos\th & 1-\cos\th\cr 1-\cos\th & 1+\cos\th},
\\
Q&=&\frac{1}{2}\pmatrix{1+\cos^2\th & 1-\cos^2\th\cr 1-\cos^2\th & 1+\cos^2\th}.
\eeqa
We notice that the matrices $U$ and $R$ are respectively orthogonal and symmetric.
We have finally
\beq
T=1+\cos^2\th.
\label{thalf}
\eeq

\subsection{The general case}

In the general case ($S$ arbitrary),
the eigenvalues of $\H$ are $E_n=n=-S,\dots,S$, again with unit steps.
The matrix $U$ encoding the corresponding eigenstates in the preferred basis
is nothing but the Wigner `small' rotation matrix $\ds(\th)$,
describing the transformation of a spin $S$ under a rotation
of angle $\th$ around the $y$-axis.
References~\cite{mess,feyn,edmonds,rose} give a full account of its properties.
The rotation matrix $\ds(\th)$ is a real orthogonal matrix,
whose entries obey the symmetry property
\beq
\ds_{a,n}(\th)=(-1)^{a-n}\ds_{-a,-n}(\th).
\label{dsym}
\eeq
This identity allows us to recast the entries of the matrix $R$ as
\beq
R_{an}=(-1)^{a-n}\ds_{a,n}(\th)\ds_{-a,-n}(\th).
\eeq
Expanding the product of rotation matrix elements
over irreducible representations~\cite{mess,feyn,edmonds,rose}, we obtain
\beq
R_{an}=\sum_{j=0}^{N-1}\phij_a\phij_nP_j(\cos\th).
\label{rexp}
\eeq
In this expression,
the integer $j$ takes $N=2S+1$ values,
\beq
P_j(\cos\th)=\dj_{0,0}(\th)
\eeq
are the Legendre polynomials, and
\beq
\phij_n=(-1)^n\braket{S,n,S,-n}{j,0}
\eeq
are Clebsch-Gordan coefficients, up to signs.

The identity~(\ref{rexp}) has several consequences.
First, it shows that $R$ is a real symmetric matrix.
The properties that $U$ and $R$ are respectively orthogonal and symmetric,
already observed for $S=\half$, thus hold for an arbitrary spin~$S$.
Furthermore,~(\ref{rexp}) yields the spectral decomposition of the matrix $R$:
$r_j=P_j(\cos\th)$ are its eigenvalues,
with the corresponding normalized eigenvectors $\phij_n$
being independent of the angle $\th$.
Finally, the matrix $R$ being symmetric,~(\ref{rrt}) implies $Q=R^2$,
and so the eigenvalues of~$Q$ are $q_j^{\vphantom{2}}=r_j^2$.
We thus obtain the explicit formula
\beq
T=\sum_{j=0}^{N-1}P_j^2(\cos\th).
\label{tp}
\eeq

Here are a few immediate properties of the above expression.
First, it is invariant when changing $\th$ to $\pi-\th$, as should be.
The Legendre polynomials indeed obey the symmetry property
$P_j(-x)=(-1)^jP_j(x)$, so that $T$ is an even function of $x=\cos\th$.
For $\th=0$ or $\th=\pi$, we have exactly $T=N$.
The Hamiltonian indeed reads $\H=\pm S_z$,
and so its eigenstates coincide with the preferred basis.
This is one example of the situation referred to in section~\ref{reminder}
as {\it no equilibration}.
For $S=\half$, i.e., $N=2$,
the result~(\ref{thalf}) is recovered, as $P_0(x)=1$ and $P_1(x)=x$.

\subsection{Behavior at large $N$ (semi-classical regime)}

The goal of this section is to estimate the growth law of $T$
in the semi-classical regime of large values of $N$.
The quantity $T$ is clearly an increasing function of~$N$,
at fixed angle $\th$,
as the expression~(\ref{tp}) is a sum of positive terms.

Consider first the mean value $\mean{T}$,
assuming that the magnetic field~${\bf h}$ acting on the spin
is isotropically distributed.
This amounts to $x=\cos\th$ being uniformly distributed in the range $-1\le x\le 1$.
The identity (see e.g.~\cite[Vol.~II, Ch.~10]{htf})
\beq
\int_{-1}^1P_j^2(x)\,\dd x=\frac{2}{2j+1}
\eeq
yields
\beq
\mean{T}=\sum_{j=0}^{N-1}\frac{1}{2j+1}=H_{2N}-\half H_N,
\eeq
where $H_N$ are the harmonic numbers, and hence
\beq
\mean{T}\approx\frac{1}{2}\ln\,(4\e^\gamma N),
\label{tlave}
\eeq
where $\gamma=0.577\,215\dots$ is Euler's constant.
The mean value $\mean{T}$ therefore grows logarithmically with $N$.

\begin{figure}[!ht]
\begin{center}
\includegraphics[angle=-90,width=.6\linewidth]{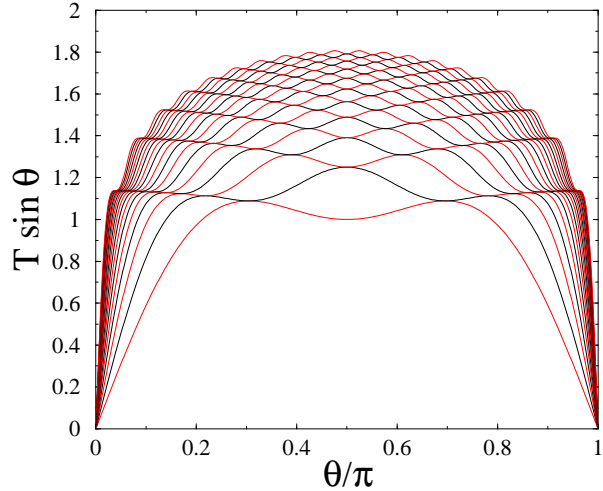}
\caption{\small
Product $T\sin\th$ against $\th/\pi$
for $N$ ranging from 2 to 20 (bottom to top).
Black curves: integer spin $S$ ($N$ odd).
Red curves: half-integer spin $S$ ($N$ even).
The lowest curve corresponds to $S=\half$ (see~(\ref{thalf})).}
\label{sint}
\end{center}
\end{figure}

A similar logarithmic growth holds at any fixed value of the tilt angle $\th$.
This can be shown by means of the generating series~\cite{bai,max,zud}
\beq
\sum_{j\ge0}P_j^2(\cos\th)\,z^j=\frac{2}{\pi(1-z)}\,\K(k),\qquad
k^2=-\frac{4z\sin^2\th}{(1-z)^2},
\eeq
where $\K(k)$ is the complete elliptic integral of the first kind.
We are interested in the singular behavior of the above expression as $z\to1$.
As the modulus~$k$ diverges in that limit,
it is advantageous to change the square modulus~\cite[Vol.~II, Ch.~13]{htf} from $k^2$ to
\beq
q^2=-\frac{k^2}{1-k^2}=\frac{4z\sin^2\th}{(1-z)^2+4z\sin^2\th}.
\eeq
We have then $\K(k)=q'\,\K(q)$, with $q'=\sqrt{1-q^2}$.
Furthermore, as $q\to1$, i.e., $q'\to0$,
the new elliptic integral diverges as $\K(q)\approx\ln\,(4/q')$,
so that
\beq
\sum_{j\ge0}P_j(\cos\th)^2\,z^j\approx\frac{1}{\pi\sin\th}\ln\frac{8\sin\th}{1-z},
\eeq
and finally
\beq
T\approx\frac{1}{\pi\sin\th}\ln\,(8\e^\gamma N\sin\th).
\label{tlth}
\eeq
The quantity $T$ thus grows logarithmically in $N$
for any fixed value of the angle $\th$,
with an angle-dependent amplitude $1/(\pi\sin\th)$.
Hence the product $T\sin\th$,
plotted in figure~\ref{sint} against the reduced angle $\th/\pi$,
grows to leading order as $(1/\pi)\ln N$, irrespective of $\th$.
Finally, the growth law~(\ref{tlave}) of the mean value $\mean{T}$
can be recovered by integrating~(\ref{tlth}) over $0\le\th\le\pi$
with the measure $\half\sin\th\,\dd\th$.

\subsection{Scaling behavior at small angle}

The quantity $T$ obeys a non-trivial scaling behavior in the regime where
$N$ is large and the tilt angle $\th$ is small,
interpolating between the growth laws
$T=T_\max=N$ for $\th=0$ and $T\sim\ln N$ for any non-zero~$\th$ (see~(\ref{tlth})).

This scaling behavior is inherited from the scaling formula
obeyed by the Legendre polynomials for $j$ large and $\th$ small,
sometimes referred to as Hilb's formula
(see e.g.~\cite[Vol.~II, Ch.~10]{htf}),
i.e.,
\beq
P_j(\cos\th)\approx J_0(s),
\label{hilb}
\eeq
with
\beq
s=j\th
\eeq
and $J_0$ being the Bessel function.
Inserting this into~(\ref{tp}), we obtain the scaling law
\beq
T\approx NF(u),
\label{tsca}
\eeq
or, equivalently,
\beq
T\th\approx G(u),
\label{tthsca}
\eeq
with
\beq
u=N\th
\eeq
and
\beq
G(u)=uF(u)=\int_0^uJ_0^2(s)\,\dd s.
\eeq

The formula~(\ref{tsca}) is more adapted to small values of the scaling variable $u$.
The function $F(u)$ decreases monotonically from $F(0)=1$ to zero.
For small $u$, the expansion
\beq
F(u)=1-\frac{u^2}{6}+\cdots
\eeq
matches the exact small-angle expansion
\beq
T=N-\frac{N(N^2-1)}{6}\,\th^2+\cdots
\eeq
of the full expression~(\ref{tp}).
We have $F(u)=1/2$ for $u\approx2.277$,
so that $T$ drops from its maximal value $T=N$ at $\th=0$
to $T=N/2$ for $\th\approx2.277/N$.

The formula~(\ref{tthsca}) is more adapted to large values of $u$.
The function $G(u)$ is monotonically increasing.
It exhibits an infinite sequence of `pauses'
at the zeros $j_n$ of the Bessel function~$J_0$,
with heights $G(j_n)=G_n$,
around which it varies cubically, as $G(u)-G_n\sim(u-j_n)^3$.
The asymptotic behavior of the Bessel function
\beq
J_0(s)\approx\left(\frac{2}{\pi s}\right)^{1/2}\cos\left(s-\frac{\pi}{4}\right)
\qquad(s\to\infty)
\eeq
implies that $G(u)$ grows logarithmically for large $u$, as
\beq
G(u)\approx\frac{1}{\pi}\ln\,(8\e^\gamma u),
\eeq
where the finite part has been obtained by matching with~(\ref{tlth}).
The zeros read approximately $j_n\approx n\pi$,
so that the heights of the pauses also grow logarithmically, as
\beq
G_n\approx\frac{1}{\pi}\ln\,(8\pi\e^\gamma n).
\eeq
There are 12 pauses with height less than 2, and 277 pauses with height less than~3.
The first of these pauses predict the heights of the ripples which are visible
on the data for finite values of $N$, shown in figure~\ref{sint}.

\begin{figure}[!ht]
\begin{center}
\includegraphics[angle=-90,width=.475\linewidth]{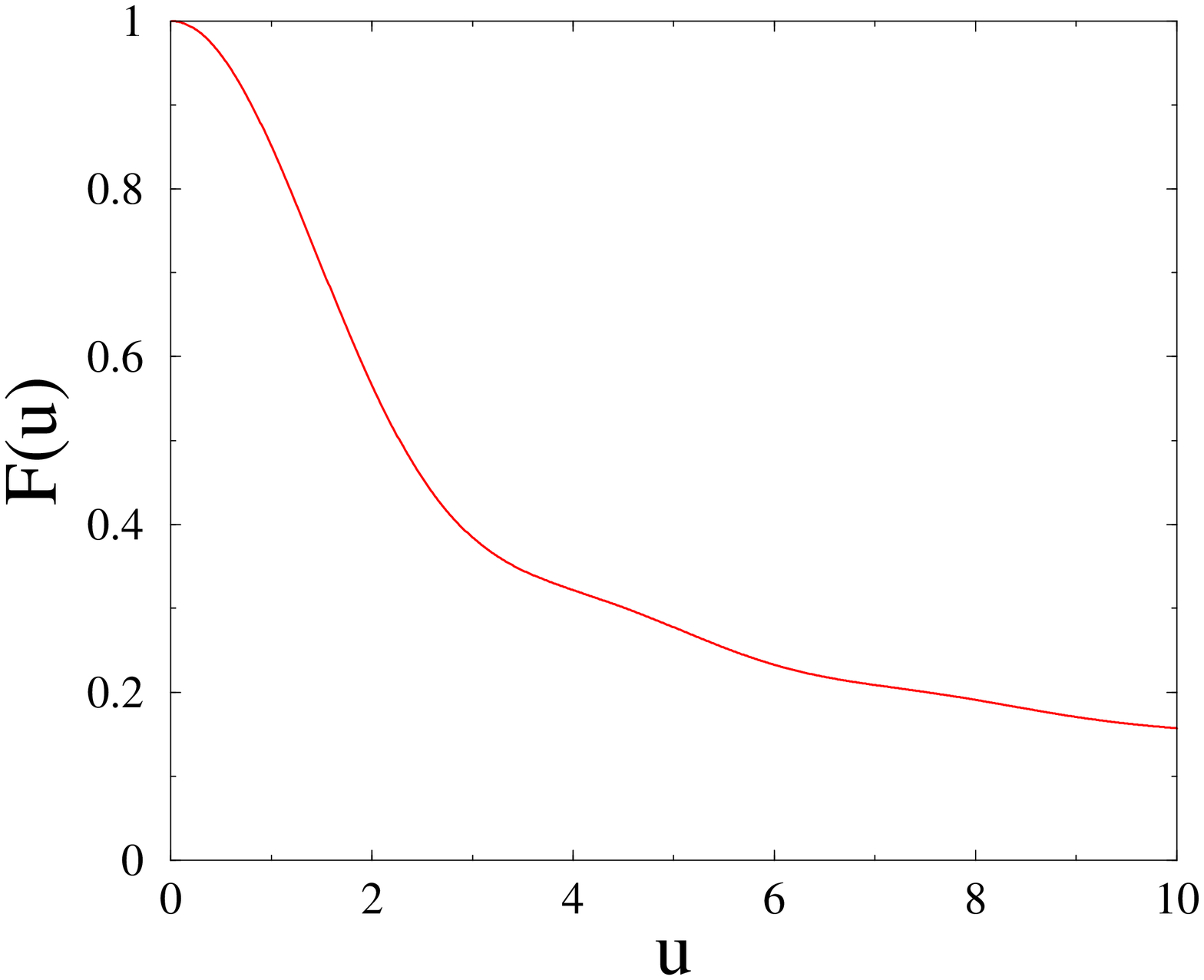}
\includegraphics[angle=-90,width=.475\linewidth]{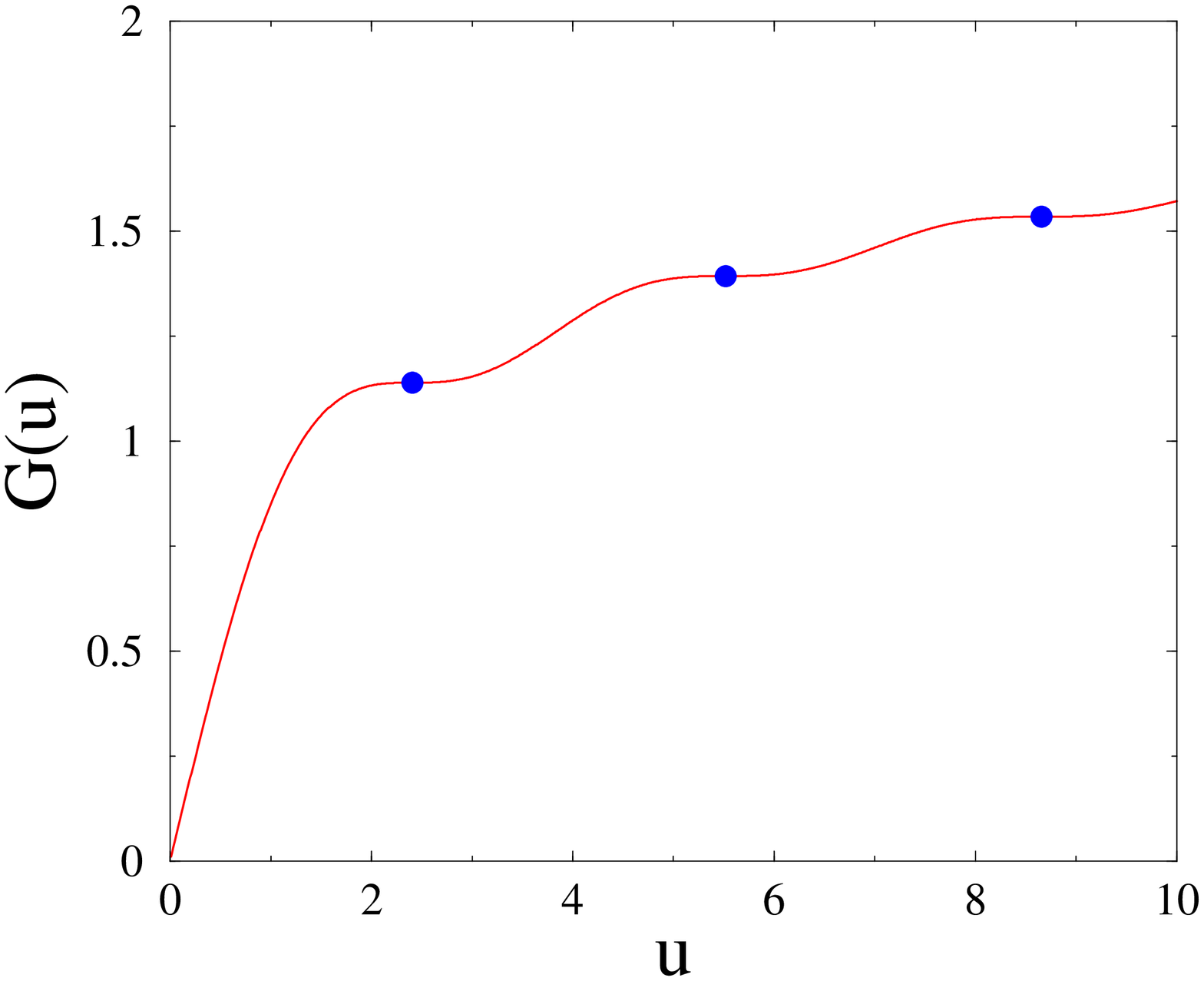}
\caption{\small
The functions $F(u)$ (left) and $G(u)=uF(u)$ (right),
respectively entering the scaling laws~(\ref{tsca}) and~(\ref{tthsca}),
against $u=N\th$.
Blue symbols on the right panel:
first three pauses
at heights $G_1\approx1.13861$, $G_2\approx1.39251$, $G_3\approx1.53384$ (see text).}
\label{fgu}
\end{center}
\end{figure}

Figure~\ref{fgu} shows the above scaling functions.
The left panel shows the function $F(u)$ entering~(\ref{tsca}).
The right panel shows the function $G(u)=uF(u)$ entering~(\ref{tthsca}).
Blue symbols show the first three pauses.

\section{A tight-binding particle on an electrified chain}
\label{stark}

In this section we investigate a tight-binding particle
on a finite chain of $N$ sites subjected to a uniform electric field $F$.
We assume $F>0$ for definiteness.

The preferential basis consists of the Wannier states~$\ket{a}$
associated with the sites of the chain ($a=1,\dots,N$).
In terms of the amplitudes $\psi_a=\braket{a}{\psi}$,
the eigenvalue equation $\H\psi=E\psi$ reads
\beq
\psi_{a+1}+\psi_{a-1}+\left(a-\half(N+1)\right)F\psi_a=E\psi_a,
\label{tbe}
\eeq
in reduced units,
where the dimensionless electric field $F$
equals the potential drop across one unit cell, in units of the hopping rate.

We shall consider an open chain with Dirichlet boundary conditions ($\psi_0=\psi_{N+1}=0$).
The choice of the constant offset in the electrostatic potential
\beq
V_a=\left(\half(N+1)-a\right)F
\label{pot}
\eeq
ensures that the latter vanishes in the middle of the sample.
As a consequence, the energy spectrum is symmetric with respect to $E=0$.
Indeed, if the wavefunction~$\psi_a$ describes a state with energy $E$,
then $(-1)^a\psi_{N+1-a}$ describes a state with energy $-E$.
Both states have the same IPR.
In particular, samples with odd sizes $N$ have a single non-degenerate eigenstate at $E=0$.

The problem of an electron in a uniform electric field
is an old classic of Quantum Mechanics.
The spectrum of a tight-binding particle on the infinite chain
has been known for long~\cite{khm,pf,ho}.
It consists of an infinite sequence of localized states,
the corresponding energies having a constant spacing
($\Delta E=F$ in our reduced units).
Such a discrete spectrum has been dubbed a Wannier-Stark ladder~\cite{w1,w2,w3}.
The fate of these ladders on finite samples has been investigated
by several authors in the 1970s~\cite{hj,sg,ms,fb,oe,bo}.
In order to investigate the behavior of $T$
in the various regimes of system size $N$ and applied electric field $F$,
we have to thoroughly revisit these works.

\subsection{General approach}
\label{general}

A general approach consists in solving~(\ref{tbe}) on the half-line ($a\ge0$),
for an arbitrary value of the energy $E$,
with boundary conditions $\psi_0=0$ and $\psi_1=1$.
This approach is known under the name of `solving the Cauchy problem'.
Equation~(\ref{tbe}) readily yields
\beqa
\psi_2&=&E+\half(N-1)F,
\nonumber\\
\psi_3&=&\left(E+\half(N-1)F\right)\left(E+\half(N-3)F\right)-1,
\label{psi23}
\eeqa
and so on.
The amplitude $\psi_a=P_{a-1}(E)$ thus obtained is a polynomial with degree $a-1$ in $E$,
dubbed a Lommel polynomial~\cite{sg}.
It also depends parametrically on $N$ and~$F$.

On a finite chain of $N$ sites with Dirichlet boundary conditions,
the energy levels~$E_n$ are the $N$ zeros of the polynomial equation
\beq
\psi_{N+1}=P_N(E)=0,
\label{pole}
\eeq
whereas the IPR $I$ of the eigenstate associated with energy $E$
reads
\beq
I=\frac{\sum_{a=1}^N\psi_a^4}{\left(\sum_{a=1}^N\psi_a^2\right)^2},
\label{irat}
\eeq
in terms of the above amplitudes $\psi_a=P_{a-1}(E)$.

It is useful to begin with a detailed study of the simplest case, namely $N=2$.
Equation~(\ref{pole}) giving the two energy levels reads
\beq
(E+\half F)(E-\half F)-1=0.
\label{ep1}
\eeq
Inserting the expressions~(\ref{psi23}) into~(\ref{irat}), we obtain
\beq
I=\frac{1+(E+\half F)^4}{(1+(E+\half F)^2)^2}.
\label{ep2}
\eeq
Eliminating the energy $E$ between~(\ref{ep1}) and~(\ref{ep2}), we obtain
\beq
AI^2+BI+C=0,
\label{ep3}
\eeq
with
\beq
A=(4+F^2)^2,\quad B=-2(4+F^2)(2+F^2),\quad C=(2+F^2)^2.
\eeq
To sum up, the two energy levels $E_n$ are the zeros of~(\ref{ep1}),
while the corresponding IPR $I_n$ are the zeros of~(\ref{ep3}).
The quantity $T$ reads (see~(\ref{tdef}))
\beq
T=I_1+I_2=-\frac{B}{A}=\frac{2(2+F^2)}{4+F^2}.
\eeq
In the present situation, both IPR are equal ($I_1=I_2$),
and so the polynomial entering~(\ref{ep3}) is a perfect square.
This peculiarity however plays no role in the exposition of the method.

The pattern for an arbitrary size $N$ is as follows.
The energy levels~$E_n$ are the zeros of the polynomial equation~(\ref{pole}),
while~(\ref{irat}) allows one to express the IPR $I$
as a rational function of the energy $E$.
As a consequence, the $I_n$ are the zeros of a polynomial equation of degree $N$,
of the form
\beq
\Pi_N(I)=\sum_{k=0}^N\Pi_{N,k}I^k=0,
\eeq
where $\Pi_N(I)$ is the resultant of the algebraic elimination of $E$
between~(\ref{pole}) and~(\ref{irat}).
This observation simplifies the calculation of $T$ for all system sizes $N$.
This quantity is indeed the sum of the $I_n$ (see~(\ref{tdef})), and so
\beq
T=-\frac{\Pi_{N,N-1}}{\Pi_{N,N}}.
\eeq

This procedure can be easily automated and implemented in a symbolic algebra software.
We thus obtain the following expressions for $T$.

\subsubsection*{$\bullet$ $N=2:$}

\beq
T=\frac{2(2+F^2)}{4+F^2}.
\label{t2}
\eeq

\subsubsection*{$\bullet$ $N=3:$}

\beq
T=\frac{5+4F^2+3F^4}{(2+F^2)^2}.
\label{t3}
\eeq

\subsubsection*{$\bullet$ $N=4:$}

\beq
T=\frac{4(24+107F^2+91F^4+48F^6+36F^8)}{(1+2F^2)(5+2F^2)(16+12F^2+9F^4)}.
\label{t4}
\eeq

\subsubsection*{$\bullet$ $N=5:$}

\beq
T=\frac{{\cal N}_5(F)}{(3+4F^2+4F^4)^2(4+24F^2+9F^4)},
\label{t5}
\eeq
with
\beqa
{\cal N}_5(F)=2(24&+&206F^2+615F^4+1676F^6
\nonumber\\
&+&700F^8+528F^{10}+360F^{12}).
\eeqa

\medskip

The following observations can be made at this stage.
The quantity $T$ is a rational even function of the applied field $F$,
whose complexity grows rapidly with $N$.
Its degree in $F$ reads $d_N=N^2/2$ if $N$ is even,
and $d_N=(N^2-1)/2$ if $N$ is odd.
We recall that the degree of a rational function
is the larger of the degrees of its numerator and denominator.
In the present situation, these degrees coincide.

In the small-field limit, we have
\beq
T=T^{(0)}+T^{(2)}F^2+\cdots
\label{tpert}
\eeq
The first term is the known value of $T$ for a free tight-binding particle~\cite{I}, i.e.,
\beq
T^{(0)}=\left\{\matrix{
\frad{3N}{2(N+1)}\quad&(N\hbox{ even}),\cr\cr
\frad{3N+1}{2(N+1)}\quad&(N\hbox{ odd}),
}\right.
\label{tfree}
\eeq
which saturates to the finite limit
\beq
T^{(0)}=\frac{3}{2}
\label{tasyfree}
\eeq
for large system sizes.
The coefficient $T^{(2)}$
will be calculated in section~\ref{pert} for all $N$ by means of perturbation theory
(see~(\ref{teven}),~(\ref{todd})).

In the large-field limit, $T$ obeys a linear-plus-constant growth of the form
\beq
T\approx NI(F)+J(F),
\label{tijlarge}
\eeq
asserting the poor equilibration properties characteristic of a localized regime, with
\beqa
I(F)&=&1-\frac{4}{F^2}+\frac{9}{F^4}-\frac{100}{9F^6}+\frac{1225}{144F^8}+\cdots,
\label{if}
\\
J(F)&=&\frac{4}{F^2}-\frac{2}{F^4}-\frac{104}{3F^6}+\frac{15943}{108F^8}+\cdots
\label{jf}
\eeqa
As $N$ increases,
more and more terms of the above asymptotic expansions stabilize.
This localized regime will be further investigated in sections~\ref{global}
and~\ref{bessel}.

\subsection{Perturbation theory}
\label{pert}

The goal of this section is to determine the coefficient $T^{(2)}$
of the expansion~(\ref{tpert}) by means of second-order perturbation theory.
The derivation follows the lines of section~5.1 of Reference~\cite{I},
devoted to the case of a weak diagonal random potential.

In the absence of an electric field,
the energy eigenvalues of a free tight-binding particle~\cite{I} read
\beq
E_n=2\cos\frac{n\pi}{N+1}
\label{en}
\eeq
($n=1,\dots,N$).
The corresponding normalized eigenstates are
\beq
\braket{a}{n}=\sqrt\frac{2}{N+1}\,\sin\frac{an\pi}{N+1}.
\label{an}
\eeq
Their IPR read
\beq
I_n=\frac{1}{2(N+1)}\left(3+\delta_{n,\half(N+1)}\right).
\label{iprfree}
\eeq
For odd sizes $N$,
the IPR of the central eigenstate at zero energy,
with number $n=\half(N+1)$, is enhanced by a factor $4/3$.
The expression~(\ref{tfree}) can be recovered by summing~(\ref{iprfree}) over $n$.

In the presence of an electric field, the un-normalized $n$th eigenstate reads
\beq
\ket{n}_F=\ket{n}-F\sum_mB_{nm}\ket{m}+F^2\sum_{m\ne n}C_{nm}\ket{m}+\cdots,
\label{nf}
\eeq
with
\beq
B_{nm}=\frac{\braopket{m}{W}{n}}{E_n-E_m},\quad
C_{nm}=\frac{1}{E_n-E_m}\sum_l\frac{\braopket{m}{W}{l}\braopket{l}{W}{n}}{E_n-E_l},
\eeq
where $E_n$ and $\ket{n}$ are given by~(\ref{en}) and~(\ref{an}),
while $W$ is the operator
\beq
W=\sum_a\left(\half(N+1)-a\right)\ket{a}\bra{a},
\eeq
where the quantity in the parentheses is the ratio $V_a/F$ (see~(\ref{pot})).
The property $\braopket{n}{W}{n}=0$ has been used in the derivation of~(\ref{nf}).

The expansion~(\ref{nf}) yields the following expansions to order $F^2$:
\beqa
\sum_a\braket{a}{n}_F^2&=&1+F^2\sum_mB_{nm}^2+\cdots,
\nonumber\\
\sum_a\braket{a}{n}_F^4&=&T^{(0)}
+6F^2\sum_a\braket{a}{n}^2\Bigl(\sum_mB_{nm}\braket{a}{m}\Bigr)^2
\nonumber\\
&+&4F^2\sum_a\braket{a}{n}^3\sum_{m\ne n}C_{nm}\braket{a}{m}+\cdots
\eeqa
We thus obtain the following expression for the coefficient $T^{(2)}$:
\beqa
T^{(2)}&=&6\sum_{a,n}\braket{a}{n}^2\Bigl(\sum_mB_{nm}\braket{a}{m}\Bigr)^2
\nonumber\\
&+&4\sum_{a,n}\braket{a}{n}^3\sum_{m\ne n}C_{nm}\braket{a}{m}
-2\sum_{a,n}\braket{a}{n}^4\sum_mB_{nm}^2.
\label{tint}
\eeqa

The final step of the derivation, i.e., the evaluation of the above sums,
could not be done by analytical means.
We have instead used the following trick.
We know from the general approach of section~\ref{general}
that $T^{(2)}$ is a rational number.
A numerical evaluation of~(\ref{tint}) with very high accuracy
($N=400$ can be reached in one minute of CPU time)
reveals that the denominator of $T^{(2)}$ only grows linearly with the system size.
It is indeed observed to be a divisor of $140(N+1)$ if $N$ is even,
and of $105(N+1)$ if $N$ is odd.
This serendipitous situation allows one to unambiguously reconstruct the numerator
of $T^{(2)}$ by means of a polynomial fit.
We are thus left with the exact closed-form expressions:

\subsubsection*{$\bullet$ $N$ even:}

\beq
T^{(2)}=\frad{N(N+2)(2N^4+8N^3+91N^2+166N+153)}{10080(N+1)}.
\label{teven}
\eeq

\subsubsection*{$\bullet$ $N$ odd:}

\beq
T^{(2)}=\frad{(N-1)(N+3)(4N^4+16N^3-121N^2-274N-525)}{20160(N+1)}.
\label{todd}
\eeq

\medskip

Up to $N=5$, the above results agree with the expansions
of the exact expressions~(\ref{t2})--(\ref{t5}).
The coefficient $T^{(2)}$ turns out to be negative for $N=3$ and $N=5$,
resulting in a non-monotonic dependence of $T$ on the applied field $F$.

For large system sizes, the expansion~(\ref{tpert}) becomes
\beq
T\approx\frac{3}{2}+\frac{N^5F^2}{5040}+\cdots
\label{tpres}
\eeq
We shall come back to this result in section~\ref{airy} (see~(\ref{philow})).

\subsection{Wannier-Stark spectrum on the infinite chain}
\label{wsladder}

The problem of a tight-binding particle on the infinite electrified chain
has been studied long ago~\cite{khm,pf,ho,w1,w2,w3}.
The energy spectrum is an infinite Wannier-Stark ladder of the form $E_k=kF$,
forgetting about the constant offset $(N+1)/2$ in~(\ref{tbe}).

\subsubsection{Spatial structure of Wannier-Stark states.}

The central Wannier-Stark state with energy
\beq
E_0=0
\eeq
is described by the normalized wavefunction
\beq
\psi_a=J_a(2/F),
\label{psia}
\eeq
where $J_a$ is the Bessel function.
We have $\psi_{-a}=(-1)^a\psi_a$,
so that the probabilities~$\psi_a^2$ are symmetric around the origin.
All the other states are obtained from the above central one
by translations along the chain.
The state with energy
\beq
E_k=kF
\eeq
is described by the wavefunction
\beq
\psi_a=J_{a-k}(2/F),
\label{psiws}
\eeq
centered at site $k$.

There is a formal identity between the wavefunction~(\ref{psia})
describing the central stationary Wannier-Stark state on the electrified chain
and the time-dependent wavefunction~\cite{toro,klm}
\beq
\psi_a(t)=\ii^{-a}J_a(2t),
\label{psit}
\eeq
describing the ballistic spreading of a quantum walker
modelled as a free tight-binding particle launched from the origin at time $t=0$,
with time $t$ playing the role of the inverse field~$1/F$,
in our reduced units.

\begin{figure}[!ht]
\begin{center}
\includegraphics[angle=-90,width=.6\linewidth]{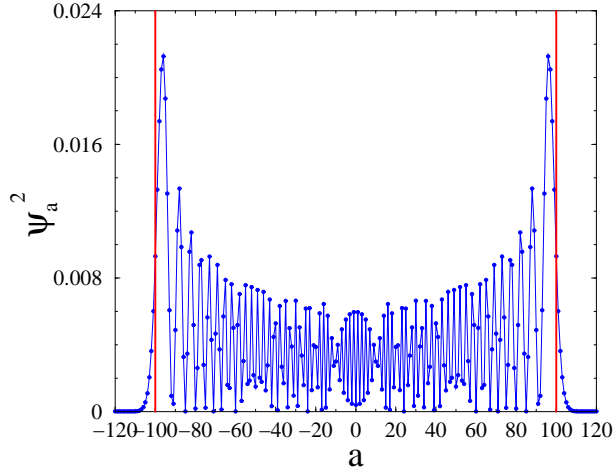}
\caption{\small
Square wavefunction $\psi_a^2$ of the central Wannier-Stark state
against position $a$, for an electric field $F=0.02$.
Vertical lines show the nominal positions of the peaks in the transition
regions ($a=\pm100$).}
\label{ladder}
\end{center}
\end{figure}

The regime of most physical interest is that of a small electric field.
The square wavefunction of the central Wannier-Stark state
is shown in figure~\ref{ladder} for $F=0.02$.
We summarize here for future use its behavior in various regions in the small-field regime.
The following estimates stemming from the theory of Bessel functions~\cite{htf,watson}
can be recovered by semi-classical techniques based on the saddle-point method,
as described in the physics literature~\cite{toro,klm}.

\subsubsection*{Allowed region ($\abs{aF}<2$).}

The wavefunction only takes appreciable values in an allowed region
extending up to a distance $2/F$ on either side of the origin.
The size of the allowed region,
\beq
L(F)=4/F,
\label{ell}
\eeq
plays the role of a localization length.
More precisely, setting
\beq
aF=2\cos\th\qquad(0<\th<\pi),
\label{aallowed}
\eeq
we have
\beq
\psi_a\approx\left(\frac{F}{\pi\sin\th}\right)^{1/2}
\cos\left(\frac{2}{F}\,(\sin\th-\th\cos\th)-\frac{\pi}{4}\right).
\label{jallowed}
\eeq
The allowed region is limited by two transition regions,
located symmetrically near $a=\pm L(F)/2=\pm2/F$,
corresponding to $\th\to0$ and $\th\to\pi$.

\subsubsection*{Transition regions ($\abs{aF}\approx2$).}

In the transition regions at the endpoints of the allowed region,
the wavefunction exhibit sharp peaks.
In the right transition region, setting
\beq
aF=2+zF^{2/3},
\label{atrans}
\eeq
we have
\beq
\psi_a\approx F^{1/3}\Ai(z),
\label{jtrans}
\eeq
where $\Ai(z)$ is the Airy function.
The width of the transition regions,
\beq
\lambda(F)\sim F^{-1/3},
\label{lam}
\eeq
provides a second length scale which diverges in the small-field regime,
albeit much less rapidly that $L(F)$ (see~(\ref{ell})).

\subsubsection*{Forbidden regions ($\abs{aF}>2$).}

The wavefunction falls off essentially exponentially fast in the forbidden regions.
In the right forbidden region, setting
\beq
aF=2\cosh\mu\qquad(\mu>0),
\label{aforbidden}
\eeq
we have
\beq
\psi_a\approx\left(\frac{F}{4\pi\sinh\mu}\right)^{1/2}
\exp\left(-\frac{2}{F}\,(\mu\cosh\mu-\sinh\mu)\right).
\label{jforbidden}
\eeq

\subsubsection{IPR of Wannier-Stark states.}

All Wannier-Stark states have the common IPR
\beq
I(F)=\sum_aJ_a^4(2/F).
\label{idef}
\eeq
This quantity was already investigated~\cite{toro}
in the context of the time-dependent wavefunction~(\ref{psit}).
It has also attracted some attention in the mathematical literature
(see~\cite{sf,ll} and references therein).

In the large-field regime ($F\gg1$),
the central Wannier-Stark state is very tightly localized near the origin.
The amplitudes $\psi_a$ indeed fall off as
\beq
\psi_a\approx\frac{1}{a!\,F^a}\qquad(a>0).
\eeq
The expansion~(\ref{if}) of $I(F)$ can therefore be recovered easily.
An all-order asymp\-totic expansion can be derived by means of the following expression
of the Mellin transform~\cite{toro}:
\beqa
M(s)&=&\int_0^\infty F^{-s-1}\,I(F)\,\dd F
\nonumber\\
&=&\frac{\Gamma(s)}{2^{3s}\pi^{3/2}}\cos\frac{s\pi}{2}
\left(\frac{\Gamma\left(\frac{1-s}{2}\right)}
{\Gamma\left(\frac{2-s}{2}\right)}\right)^3\quad(0<\Re s<1).
\eeqa
Summing the contributions of the simple poles at $s=-2,-4,\dots$, we obtain
\beq
I(F)=\sum_{k\ge0}\frac{(-1)^k(2k)!^2}{k!^6\,F^{2k}}.
\eeq
The above result can be recovered by using the
expression of $I(F)$ in terms of a hypergeometric function~\cite{sf,ll}:
\beq
I(F)=\null_2{\rm F}_3\left(\frac{1}{2},\frac{1}{2};1,1,1;-\frac{16}{F^2}\right).
\eeq

In the small-field regime, we have
\beq
I(F)=\frac{F}{2\pi^2}\ln\frac{64\e^\gamma}{F}
+\frac{F^{3/2}}{16\sqrt{\pi}}\,\sin(8/F-\pi/4)+\cdots
\label{ismall}
\eeq
The first term of the above expression exhibits a logarithmic correction,
stemming from the double pole of $M(s)$ at $s=1$,
with respect to the leading scaling $I(F)\sim1/L(F)\sim F$.
The presence of this logarithmic correction to scaling
was already emphasized in the time-dependent problem~\cite{toro},
where it was put in perspective with the following phenomenon dubbed bifractality.
The generalized IPR
\beq
I^{(q)}(F)=\sum_a\abs{J_a(2/F)}^{2q},
\eeq
with $q>0$ arbitrary,
exhibits an anomalous scaling behavior as $F\to0^+$, of the form
\beq
I^{(q)}(F)\sim F^{\tau(q)},
\eeq
with
\beq
\tau(q)=\left\{\matrix{
q-1\hfill&(q<2),\hfill\cr\cr
\frad{2q-1}{3}\quad&(q>2).\hfill\cr
}\right.
\eeq
The usual IPR $I(F)$ corresponds to $q=2$,
i.e., precisely the break point between both branches
of the above bifractal spectrum, where the generalized IPR is respectively dominated
by the allowed region (for $q<2$) and by the transition regions (for $q>2$).

The subleading oscillatory contribution in~(\ref{ismall}),
whose derivation requires more sophisticated techniques~\cite{ll},
results in a non-monotonic behavior of the IPR $I(F)$ as a function of the field $F$.

\subsection{Main features of the IPR $I_n$ and of $T$}
\label{global}

In this section we provide a picture of the salient features
of the IPR $I_n$ of the various eigenstates, and of their sum $T$.

\begin{figure}[!ht]
\begin{center}
\includegraphics[angle=-90,width=.6\linewidth]{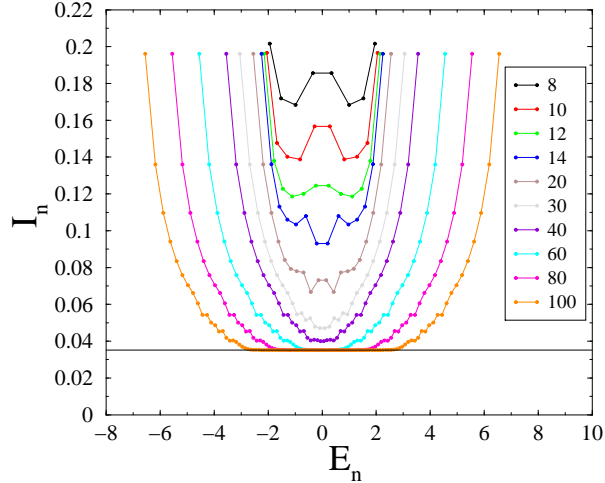}
\caption{\small
IPR $I_n$ against energy $E_n$
for all eigenstates of finite chains of various lengths $N$,
for an applied field $F=0.1$.
Horizontal line: $I(0.1)=0.035\,222\dots$}
\label{iprs}
\end{center}
\end{figure}

Figure~\ref{iprs} shows the IPR $I_n$ against the energy $E_n$
for all eigenstates of finite chains of various lengths ranging from $N=8$ to 100.
The applied field reads $F=0.1$, so that the localization length is $L(F)=40$.
The horizontal line shows the value $I(F)=0.035\,222\dots$ of the IPR
of Wannier-Stark states.
The IPR $I_n$ have been evaluated by means of a direct numerical diagonalization
of the Hamiltonian matrix~$\H$.
The symmetry of the plot with respect to $E=0$ is manifest.
As $N$ increases, the IPR of the few first (or last) eigenstates converge very fast
to well-defined limiting values.
In the small-field regime,
these limits will be shown to scale as $F^{1/3}$ (see~(\ref{inairy})).
They are therefore much larger than the IPR $I(F)$ of Wannier-Stark states,
which is proportional to $F$, with a logarithmic correction (see~(\ref{ismall})).
In the localized regime,
i.e., for sample sizes $N$ larger than the localization length $L(F)$,
there are approximately $N-L(F)$ bulk states in the central part of the energy spectrum:
\beq
\abs{E}<\half NF-2.
\label{center}
\eeq
These eigenstates are exponentially close to the Wannier-Stark states
whose allowed region is entirely contained in the sample (see e.g.~\cite{hj,sg,oe}).
Their IPR $I_n$ is therefore very close to~$I(F)$.
The remaining eigenstates are edge states,
living near either boundary of the system,
whose energies lie in the wings of the spectrum:
\beq
\half NF-2<\abs{E}<\half NF+2.
\label{wings}
\eeq
Their IPR vary in a broad range between $I(F)$ and $F^{1/3}$.
On shorter samples, whose length $N$ is smaller than the localization length $L(F)$,
the distinction between bulk and edge states looses its meaning.
All the IPR $I_n$ are larger than $I(F)$
and manifest a rather irregular dependence on the state number $n$.
This is exemplified by the data for system sizes $N=8$ to 20 in figure~\ref{iprs}.

The above picture substantiates the asymptotic linear-plus-constant growth
of $T$ with the system size~$N$, already advocated in~(\ref{tijlarge}):
\beq
T\approx NI(F)+J(F).
\label{tij}
\eeq
This estimate holds not only for large fields, but all over the localized regime,
defined by the inequality $N>L(F)$, i.e., $NF>4$.
The slope $I(F)$ is the IPR of Wannier-Stark states (see~(\ref{idef})).
The offset $J(F)$ somehow encodes the contributions of all edge states.
Its exact value is not known for arbitrary values of the field~$F$.
It can be noticed that the coefficients of its large-field expansion~(\ref{jf})
do not exhibit any obvious regularity.
The quantity $J(F)$ will be investigated in section~\ref{bessel}
in the small-field regime (see~(\ref{jest})).

\begin{figure}[!ht]
\begin{center}
\includegraphics[angle=-90,width=.6\linewidth]{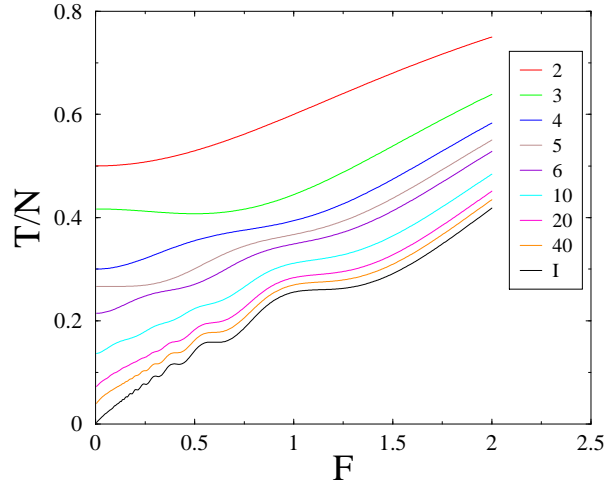}
\caption{\small
Ratio $T/N$ against $F$ for several system sizes $N$.
Lowest black line: IPR $I(F)$, corresponding to $N\to\infty$.}
\label{tn}
\end{center}
\end{figure}

Figure~\ref{tn} shows the ratio $T/N$ against $F$ for several system sizes $N$.
The lowest black line shows the IPR $I(F)$,
corresponding to the $N\to\infty$ limit of $T/N$, according to~(\ref{tij}).
The data for finite $N$ converge smoothly to that limit,
for any fixed non-zero value of the electric field $F$.

The quantitative analysis of the IPR $I_n$ and of their sum $T$
in the small-field regime will require some special care.
Two different characteristic lengths indeed diverge in the $F\to0^+$ limit,
namely $L(F)\sim1/F$ (see~(\ref{ell})) and $\lambda(F)\sim1/F^{1/3}$ (see~(\ref{lam})).
Each of these diverging lengths is responsible for the occurrence
of a non-trivial scaling regime (see sections~\ref{airy} and~\ref{bessel}).

\subsection{Airy scaling regime $(N\sim\lambda(F)\sim1/F^{1/3})$}
\label{airy}

This section is devoted to the regime where the system size $N$
and the size $\lambda(F)$ of the transition regions (see~(\ref{lam}))
are large and comparable.
We introduce the scaling variable
\beq
x=NF^{1/3},
\label{xdef}
\eeq
proportional to the ratio $N/\lambda(F)$.
Since the wavefunction of the Wannier-Stark states
in the transition regions is an Airy function (see~(\ref{jtrans})),
we dub this regime the Airy scaling regime.

\subsubsection{Extremal edge states.}

Let us first investigate the extremal edge states on a very large sample,
with energies near the left edge of the spectrum ($E\approx-\half NF-2$).
A direct approach goes as follows (see e.g.~\cite{ms,fb}).
Setting
\beq
\chi_a=(-1)^a\psi_a,
\eeq
and replacing $N+1$ by $N$ in the offset of the electrostatic potential,
the tight-binding equation~(\ref{tbe}) becomes
\beq
\chi_{a+1}+\chi_{a-1}=\left(aF-\half NF-E\right)\chi_a.
\eeq
Using a continuum description, this reads
\beq
\frac{\dd^2\chi}{\dd a^2}\approx(aF+\ecal)\chi,
\label{ddchi}
\eeq
with
\beq
\ecal=-\half NF-E-2,
\eeq
and with the boundary conditions that $\chi$ vanishes for $a=0$ and $a\to+\infty$.
Then, setting
\beq
a=(z+c)F^{-1/3},\qquad\ecal=-cF^{2/3},
\eeq
equation~(\ref{ddchi}) boils down to Airy's equation
\beq
\frac{\dd^2\chi}{\dd z^2}=z\chi,
\eeq
whose solution decaying as $z\to+\infty$ is $\chi(z)=\Ai(z)$.

The extremal edge states are quantized by the Dirichlet boundary condition at $a=0$.
The latter requires $c=c_n$,
where $-c_n$ are the zeros of the Airy function ($n=1,2,\dots$).
These zeros are real and negative, and so $c_n>0$.
The extremal edge states on a very large sample therefore have energies
\beq
E_n\approx-\half NF-2+c_n F^{2/3}.
\label{enairy}
\eeq
The IPR of these states read
\beq
I_n\approx a_n F^{1/3},
\label{inairy}
\eeq
with
\beq
a_n=\frac{\mu_4(c_n)}{\mu_2^2(c_n)}
\eeq
and
\beq
\mu_m(c)=\int_{-c}^{+\infty}\Ai^m(z)\,\dd z.
\eeq

The above results can be made more explicit for large $n$,
which turns out to be the regime of most interest.
The asymptotic behavior of the Airy function
\beq
\Ai(z)\approx\frac{1}{\sqrt{\pi}\abs{z}^{1/4}}
\sin\left(\frac{2}{3}\abs{z}^{3/2}+\frac{\pi}{4}\right)
\qquad(z\to-\infty)
\eeq
implies
\beq
c_n\approx\left(\frac{3\pi n}{2}\right)^{2/3}.
\eeq
Then, for large $c$, averaging first the powers of the rapidly varying sine function,
we obtain
\beqa
\mu_2(c)&\approx&\frac{1}{2}\int_{-c}^0\frac{\dd z}{\pi\abs{z}^{1/2}}
\approx\frac{\sqrt{c}}{\pi},
\\
\mu_4(c)&\approx&\frac{3}{8}\int_{-c}^{-c_0}\frac{\dd z}{\pi^2\abs{z}}
\approx\frac{3}{8\pi^2}\ln\frac{c}{c_0}.
\eeqa
The finite part $c_0$ of the logarithm
has been evaluated by more sophisticated tech\-niques~\cite{reid}, yielding
\beq
\mu_4(c)\approx\frac{3}{8\pi^2}\ln(4\e^{2\gamma/3}c),
\eeq
where $\gamma$ is again Euler's constant.
We thus obtain the estimate
\beq
a_n\approx\frac{1}{4}\left(\frac{3\pi n}{2}\right)^{-2/3}\ln(12\pi\e^\gamma n).
\label{anres}
\eeq

\subsubsection{Scaling behavior of $T$.}

The evaluation of $T$ for large but finite systems in the Airy scaling regime
goes as follows.
The IPR $I_n$ of the $n$th eigenstate scales as
\beq
I_n\approx\frac{1}{N}\phi_n(x),
\eeq
with
\beq
\phi_n(x)\approx\left\{\matrix{
\frad{3}{2}\hfill&(x\to0),\hfill\cr
a_nx\quad&(x\to\infty).
}\right.
\label{branches}
\eeq
The above estimates respectively match~(\ref{iprfree}) and~(\ref{inairy}).

As a consequence, $T$ obeys a scaling law
\beq
T\approx T^{(0)}+\frac{1}{N}\,\Phi(x)
\label{tfssws}
\eeq
of the very same form as~(\ref{tfss}),
corresponding to a particle in a weak disordered potential~\cite{I},
albeit with a different scaling variable $x$,
and a different function $\Phi(x)$.
The first term is the value of $T$ for a free particle,
given by~(\ref{tfree}), and so we have $\Phi(0)=0$.
The following observation~\cite{I} also applies to the present situation.
Had we chosen the limit value~(\ref{tasyfree}) for $T^{(0)}$
as the first term in~(\ref{tfssws}),
instead of the exact expression~(\ref{tfree}) for finite $N$,
the scaling function $\Phi(x)$ would have acquired
a finite offset depending on the parity of $N$,
i.e., $\Phi(0)=-3/2$ (resp.~$\Phi(0)=-1$) for $N$ even (resp.~$N$ odd).

When the scaling variable $x$ is small,
the perturbative result~(\ref{tpres}) implies
\beq
\Phi(x)\approx\frac{x^6}{5040}\qquad(x\to0).
\label{philow}
\eeq

The asymptotic behavior of the scaling function $\Phi(x)$
at large $x$ can be estimated as follows.
The crossover between both estimates entering~(\ref{branches}) remains steep,
in the sense that it does not broaden out as the state number $n$ gets larger and larger.
The sum $T$ can therefore be estimated as
\beq
T\approx\frac{2}{N}\sum_{n=1}^{N/2}\,\max\!\left(\frac{3}{2},\;a_nx\right).
\label{testimate}
\eeq
Using the expression~(\ref{anres}) for the amplitudes $a_n$,
we obtain after some algebra
\beq
\Phi(x)\approx6n_c(x),
\eeq
where $n_c(x)$ is the value of $n$
such that both arguments of the max function entering~(\ref{testimate}) coincide.
To leading order in $\ln x\gg1$, we have
\beq
n_c(x)\approx\frac{(x\ln x)^{3/2}}{12\pi},
\eeq
and so
\beq
\Phi(x)\approx\frac{(x\ln x)^{3/2}}{2\pi}\qquad(x\to\infty).
\label{highlead}
\eeq

\begin{figure}[!ht]
\begin{center}
\includegraphics[angle=-90,width=.6\linewidth]{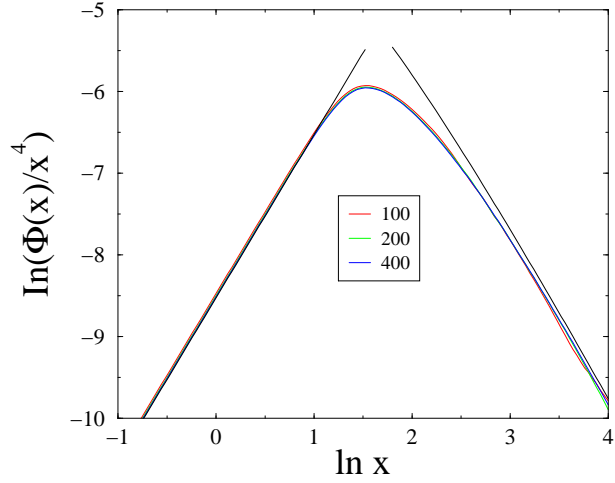}
\caption{\small
Log-log plot of the ratio $\Phi(x)/x^4$ against the scaling variable $x$.
Color lines: data pertaining to system sizes 100, 200 and 400.
Straight line to the left: power-law result~(\ref{philow}).
Line to the right: estimate~(\ref{highlead}).}
\label{sca1}
\end{center}
\end{figure}

Figure~\ref{sca1} shows a log-log plot of the ratio $\Phi(x)/x^4$
against the scaling variable~$x$.
This ratio has been chosen in order to better reveal the structure
of the scaling function~$\Phi(x)$.
Color lines show data pertaining to system sizes 100, 200 and 400.
The straight line to the left shows the power-law result~(\ref{philow})
stemming from perturbation theory.
The line to the right shows the estimate~(\ref{highlead}).
The data exhibit a very good convergence to the scaling function~$\Phi(x)$,
as well as a good agreement with both analytical predictions.
Small oscillatory corrections are however visible at very high~$x$.

\subsection{Bessel scaling regime $(N\sim L(F)\sim1/F)$}
\label{bessel}

This section is devoted to the regime where the system size $N$
and the localization length $L(F)$ (see~(\ref{ell})) are large and comparable.
We introduce the scaling variable
\beq
y=NF,
\label{ydef}
\eeq
proportional to the ratio $N/L(F)$,
and dub this regime the Bessel scaling regime.

\subsubsection{Generic edge states.}
\label{besselgeneric}

Let us first investigate the generic edge states on a very large sample
in the localized regime ($y>4$), considering the left edge for definiteness.
These edge states are very well described
by the restriction of the Wannier-stark states~(\ref{psiws}), i.e.,
\beq
\psi_a=J_{a-k}(2/F),
\label{psipart}
\eeq
to a part of their allowed region.
The allowed region of the above wavefunction is $k-2/F<a<k+2/F$.
Left edge states therefore correspond to the range $-2/F<k<2/F$,
i.e., $-\half y-2<E<-\half y+2$, in agreement with~(\ref{wings}).

The above edge states are quantized by the Dirichlet boundary condition $\psi_0=0$.
Setting
\beq
k=-\frac{2\cos\a}{F},\qquad E=-\frac{y}{2}-2\cos\a,
\eeq
the expression~(\ref{jallowed}) of the wavefunction in the allowed region
yields the quantization condition
\beq
\sin\a_n-\a_n\cos\a_n\approx\frac{\pi nF}{2}.
\label{qalpha}
\eeq
As the angle $\alpha$ increases from $0$ to $\pi$,
the edge state number $n$ ranges from 1 to $n_\edge$, where
\beq
n_\edge=\frac{2}{F}=\frac{L(F)}{2}
\eeq
is our estimate for the number of edge states in each wing of the spectrum
in the small-field regime.
The quantization condition~(\ref{qalpha}) yields the density of states
\beq
\rho(\a)=\frac{2}{\pi F}\,\a\sin\a,
\eeq
normalized as
\beq
\int_0^\pi\rho(\a)\,\dd\a=n_\edge.
\eeq

The IPR of the above edge states read
\beq
I_n=\frac{S_{4,n}}{S_{2,n}^2},
\eeq
with
\beq
S_{m,n}=\sum_{a\ge1}\psi_{a,n}^m,
\eeq
and $\psi_{a,n}$ is given by~(\ref{psipart}),
together with the quantization condition~(\ref{qalpha}).
The sums $S_{m,n}$ can be estimated by
using the expression~(\ref{jallowed}) of the wavefunction in the allowed region,
averaging first the powers of the sine function,
and transforming the sums over $a$ into integrals over $\th$
in the range $0<\th<\a_n$, according to~(\ref{aallowed}).
We thus obtain
\beqa
S_{2,n}&\approx&\frac{1}{2}\cdot\frac{2}{\pi}\int_0^{\a_n}\dd\th=\frac{\a_n}{\pi},
\\
S_{4,n}&\approx&\frac{3}{8}\cdot\frac{2F}{\pi^2}\int_\eps^{\a_n}\frac{\dd\th}{\sin\th}
=\frac{3F}{4\pi^2}\ln\left(\frac{2}{\eps}\tan\frac{\a_n}{2}\right),
\eeqa
and so
\beq
I_n\approx\frac{3F}{4\a_n^2}\ln\left(\frac{2}{\eps}\tan\frac{\a_n}{2}\right).
\label{inbessel}
\eeq
The cutoff $\eps$ is independent of the state number $n$.
It can therefore be determined by matching the Airy and Bessel regimes.
For $n\ll n_\edge$, i.e., $nF\ll1$, (\ref{qalpha}) reads
\beq
\a_n\approx\left(\frac{3\pi nF}{2}\right)^{1/3}.
\label{anbessel}
\eeq
Identifying the expressions~(\ref{inairy}),~(\ref{anres}) in the Airy regime
and~(\ref{inbessel}),~(\ref{anbessel}) in the Bessel regime yields
\beq
\eps=\left(\frac{F}{8\e^\gamma}\right)^{1/3}.
\eeq
We are thus left with the estimate
\beq
I_n\approx\ical(F,\a_n)
=\frac{F}{4\a_n^2}\ln\left(\frac{64\e^\gamma}{F}\tan^3\frac{\a_n}{2}\right).
\label{inb}
\eeq

\subsubsection{Scaling behavior of $T$.}

The evaluation of $T$ for finite systems in the Bessel scaling regime
goes as follows.
The cases $y>4$ and $y<4$ have to be dealt with separately.

\subsubsection*{$\bullet$ $y>4$.}

Consider first the simple situation of samples in the localized regime,
such that $N>L(F)$, i.e., $y>4$.
As already discussed in section~\ref{global},
there are $N-2n_\edge=N-L(F)=(y-4)/F$ bulk states,
whose IPR essentially coincide with $I(F)$,
and $2n_\edge=L(F)=4/F$ edge states,
whose IPR vary according to~(\ref{inb}).
We thus obtain
\beq
T\approx(y-4)\frac{I(F)}{F}+T_\edge,
\label{tloc}
\eeq
where the contribution of all edge states reads
\beq
T_\edge\approx2\int_0^\pi\rho(\a)\,\ical(F,\a)\,\dd\a
=A\ln\frac{64\e^\gamma}{F}+3B,
\eeq
with
\beqa
A&=&\frac{1}{\pi}\int_0^\pi\frac{\sin\a}{\a}\,\dd\a=\frac{{\rm Si}(\pi)}{\pi}
=0.589\,489\dots,
\\
B&=&\frac{1}{\pi}\int_0^\pi\frac{\sin\a}{\a}\,\ln\tan\frac{\a}{2}\,\dd\a
=-0.462\,288\dots
\eeqa

The result~(\ref{tloc}) has the expected linear-plus-constant
form~(\ref{tijlarge}),~(\ref{tij}).
It yields the following estimate for the offset $J(F)$ in the small-field regime:
\beq
J(F)\approx\left(A-\frac{2}{\pi^2}\right)\ln\frac{64\e^\gamma}{F}+3B.
\label{jest}
\eeq
Finally,~(\ref{tloc}) can be recast as
\beq
T\approx U(y)\ln\frac{1}{F}+V(y),
\label{tuv}
\eeq
with
\beqa
U(y)&=&\frac{y-4}{2\pi^2}+A,
\label{uhi}
\\
V(y)&=&\left(\frac{y-4}{2\pi^2}+A\right)\ln(64\e^\gamma)+3B.
\label{vhi}
\eeqa

\subsubsection*{$\bullet$ $y<4$.}

\begin{figure}[!ht]
\begin{center}
\includegraphics[angle=-90,width=.475\linewidth]{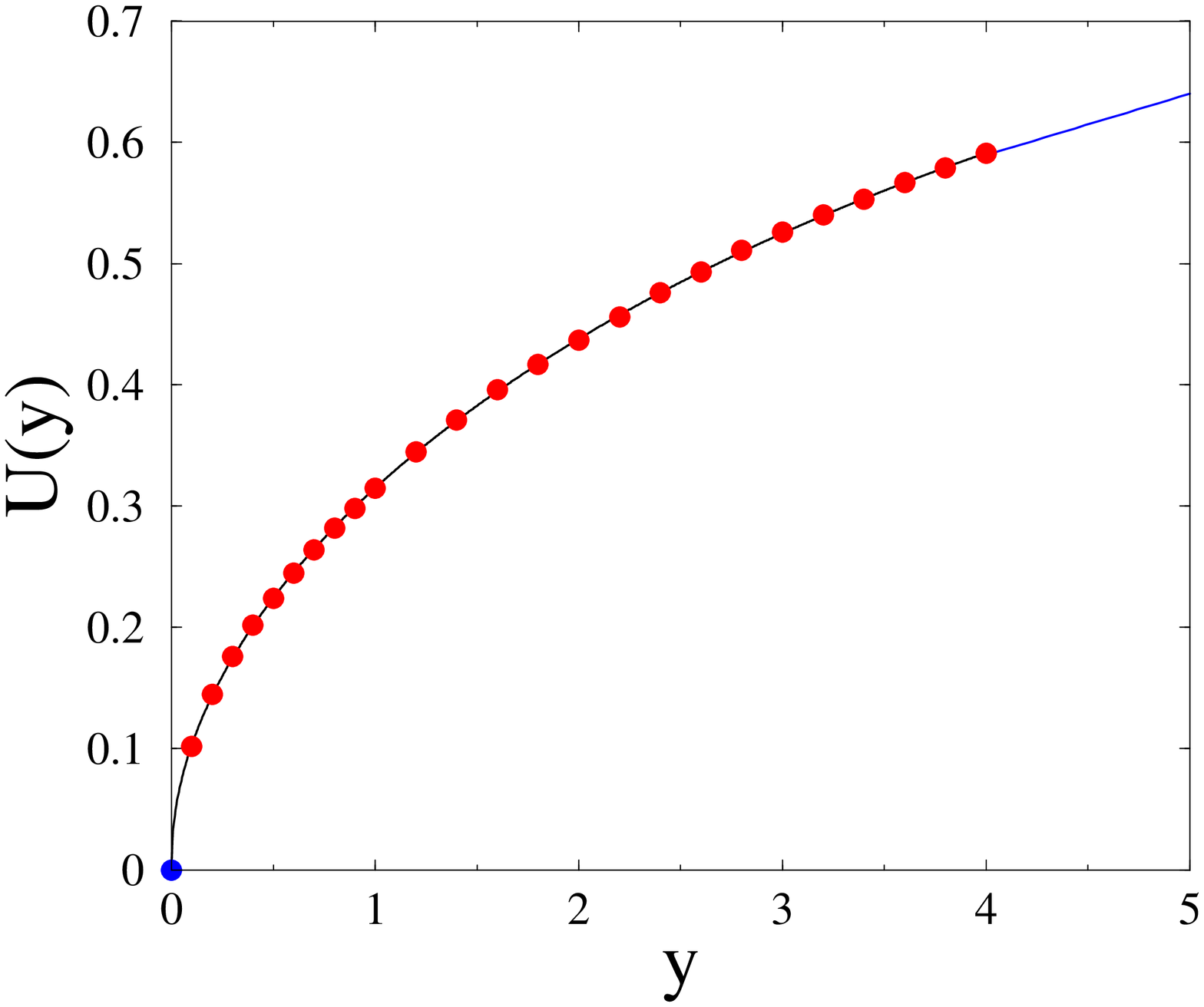}
\includegraphics[angle=-90,width=.475\linewidth]{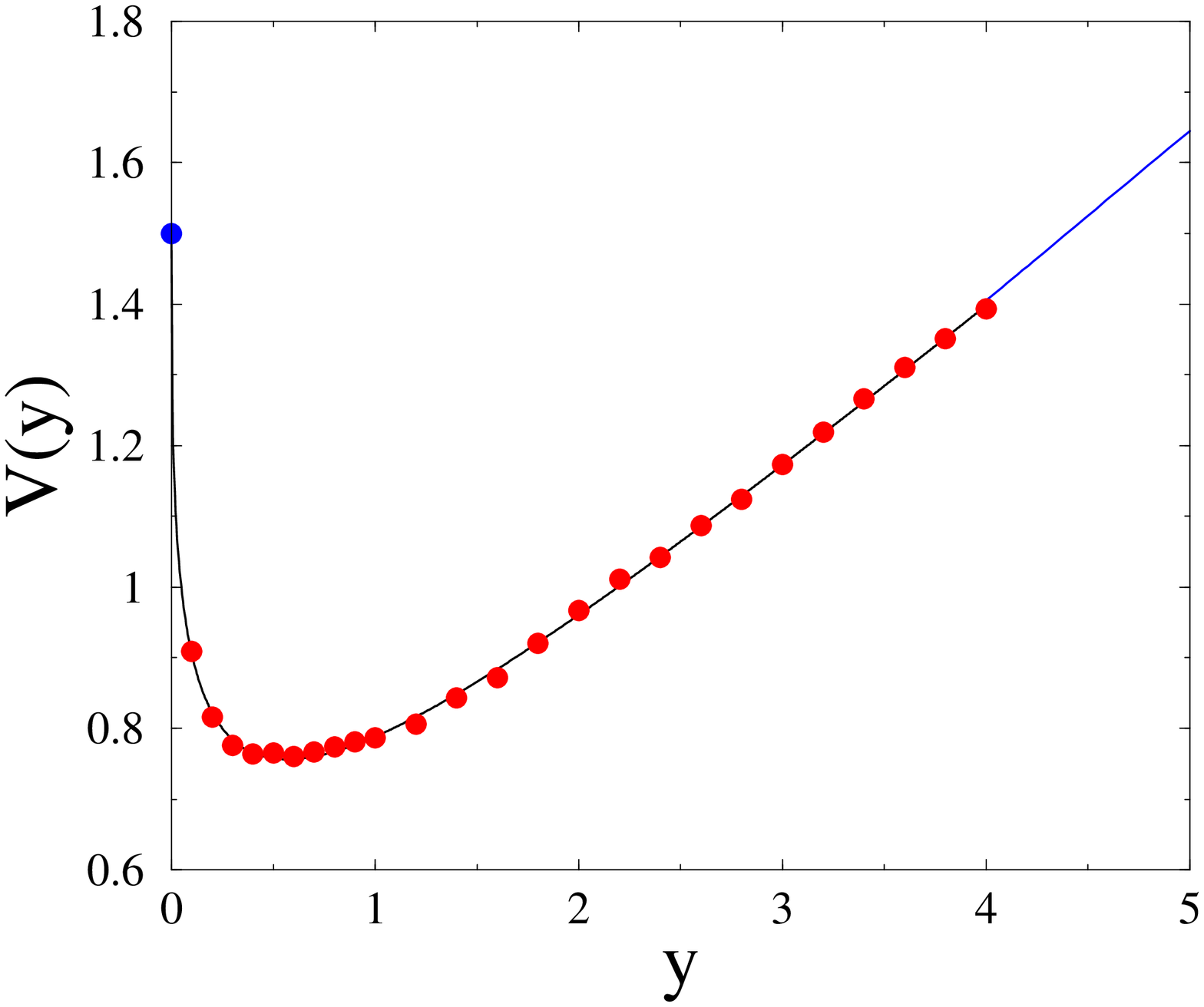}
\caption{\small
Amplitudes $U(y)$ (left) and $V(y)$ (right)
entering the scaling formula~(\ref{tuv}) for $T$.
Red symbols: numerical data for $y<4$.
Black lines: fits to these data (see text).
Blue straight lines: analytical results~(\ref{uhi}),~(\ref{vhi}) for $y>4$.
Blue symbols: values $U(0)=0$ and $V(0)=3/2$.}
\label{scauv}
\end{center}
\end{figure}

Consider now shorter samples, such that $N<L(F)$, i.e., $y<4$.
As discussed in section~\ref{global},
there is no clear-cut distinction between bulk and edge states in this situation.
As a consequence, the formula~(\ref{inb}) for the IPR of edge states
is not sufficient to derive an estimate of $T$ for $y<4$.
We nevertheless hypothesize that $T$ still obeys a logarithmic scaling law
in $F$ of the form~(\ref{tuv}),
even though the amplitudes $U(y)$ and $V(y)$ cannot be predicted analytically.
Figure~\ref{scauv} presents numerical results for these amplitudes.
For each value of $y$, $T$ has been calculated for 15
roughly equidistant system sizes $N$ on a logarithmic scale between $N=10$ and $N=5000$.
These values of $T$ obey accurately the logarithmic dependence
on $F$, i.e., on $N=y/F$, hypothesized in the scaling form~(\ref{tuv}).
The resulting estimates for $U(y)$ and $V(y)$ are shown as symbols
in both panels of figure~\ref{scauv}.
Black lines show fits of the form
$U(y)=(u_1\ln y+u_2)\sqrt{y}+(u_3\ln y+u_4)y$
and $V(y)=3/2+(v_1\ln y+v_2)\sqrt{y}+(v_3\ln y+v_4)y$.
The form of these fits is freely inspired by the expected crossover at small $y$
with the estimate~(\ref{highlead}) for large $x$ in the Airy regime, i.e.,
\beq
T\approx\frac{3}{2}+\frac{\sqrt{y}}{2\pi}\left(\ln\frac{y}{F^{2/3}}\right)^{3/2}.
\eeq
In any case, the proposed fits describe the data accurately.
Finally, the blue straight lines show
the analytical results~(\ref{uhi}),~(\ref{vhi}) for $y>4$.
The data suggest that the functions $U(y)$ and $V(y)$ are continuous at $y=4$,
as well as their first derivatives.

\section{Discussion}
\label{disc}

A novel way of investigating equilibration
-- or lack of equilibration -- in small isolated quantum systems
has been put forward recently~\cite{I}.
The key quantity of that approach is the trace $T$
of the matrix $Q$ of asymptotic transition probabilities
in some preferential basis chosen once for all.
The quantity $T$ is also the sum of the inverse participation ratios
of all eigenstates of the Hamiltonian in the same basis.
This transition matrix formalism therefore requires
the knowledge of all energy eigenvectors,
so that its scope is limited to small systems.
This is however consistent with the purpose of the approach,
which is precisely to quantify the degree of equilibration of small quantum systems.
The quantity $T$ provides a measure of the degree of equilibration of the system
if launched from a typical basis state of the preferential basis.
It may vary between $T_\min=1$ and $T_\max=N$, the dimension of the Hilbert space;
the larger $T$, the poorer the equilibration.

The models investigated in~\cite{I} and in the present work
embrace the most significant physical systems with a finite-dimensional Hilbert space.
The outcomes of both papers
allow one to get a full understanding of the behavior of the quantity $T$.
Quantum systems whose dynamics is governed by a generic Hamiltonian
exhibit good typical equilibration properties,
so that $T$ saturates to a non-trivial limit for large $N$.
This limit is however always found to be larger than $T_\min=1$.
It equals $T^{(0)}=3/2$ for a free tight-binding particle
on a large open interval~\cite{I},
and $T_\U=2$ or $T_\O=3$ for a random Hamiltonian whose distribution is invariant
under the groups U$(N)$ or O$(N)$ (section~\ref{rmt}).

An interesting situation is met when varying a system parameter
through some scaling regime
induces a continuous crossover from good to bad equilibration properties.
Three systems have been shown to exhibit such a crossover phenomenon,
where $T$ is found to obey non-trivial scaling behavior.
Table~\ref{models} summarizes the following comparative discussion of these three models.
For a spin $S$ in a tilted magnetic field (section~\ref{spin}),
making an angle $\th$ with the preferred quantization axis,
we have $T=T_\max=N$ for $\th=0$,
while $T\sim\ln N$ for any fixed non-zero tilt angle~$\th$.
Both regimes are connected by the scaling laws~(\ref{tsca}),~(\ref{tthsca}),
with scaling variable $u=N\th$.
For a tight-binding particle on a finite open chain
in a weak random potential~\cite{I},~$T$ obeys the scaling law
recalled in the Introduction~(see~(\ref{tfss})),
with scaling variable $x=Nw^{2/3}$,
which describes the crossover between the finite value $T=T^{(0)}=3/2$
of a free particle in the ballistic regime
and the linear growth $T\approx N\bar{Q}$ in the localized regime,
testifying poor equilibration,
with $\bar{Q}\approx Aw^{4/3}$ for a weak disorder.
For a tight-binding particle on a finite electrified chain (section~\ref{stark}),
$T$ obeys a similar but richer behavior
in the situation where the applied field $F$ is small,
so that the localization length $L(F)=4/F$ is large.
The quantity~$T$ indeed exhibits
two successive scaling laws describing the crossover
between the ballistic regime and the insulating one,
namely a first scaling law of the form~(\ref{tfssws})
with scaling variable $x=NF^{1/3}$ in the so-called Airy scaling regime,
followed by a second scaling law of the form~(\ref{tuv})
with scaling variable $y=NF$ in the so-called Bessel scaling regime.

\begin{table}[!ht]
\begin{center}
\begin{tabular}{|c|c|c|c|c|}
\hline
model & $N$ & control parameter & scaling variable & Ref.\\
\hline
$\dbl{spin in tilted}{magnetic field}$ & $2S+1$ & tilt angle $\th$ & $u=N\th$ & Sec.~4\\
\hline
$\dbl{particle on}{disordered chain}$ & $\dbl{number}{of sites}$ &
disorder strength $w$ & $x=Nw^{2/3}$ &\cite{I}\\
\hline
$\dbl{particle on}{electrified chain}$ & $\dbl{number}{of sites}$ &
electric field $F$ &
$\!\left\{\matrix{x=NF^{1/3}\,\mbox{(Airy)}\cr y=NF{\hskip 7.5pt}\mbox{(Bessel)}}\right.$ &
Sec.~5\\
\hline
\end{tabular}
\end{center}
\caption{\small
The three models known to exhibit
a continuous crossover from good to bad equilibration properties
in the regime where $N$ is large whereas the control parameter is small:
definition of the control parameters and of the variables entering
the scaling behavior of $T$.}
\label{models}
\end{table}

Numerical results on the XXZ spin chain~\cite{mpl}
demonstrate that a transition from good to bad equilibration properties,
in the sense of a change in the growth law of $T$,
can be observed in a clean quantum many-body system
which does not exhibit many-body localization.
There, the transition is found to coincide with a thermodynamical phase transition,
with $T\sim L$ in the gapless phase of the model ($\abs{\Delta}<1$),
and $T\sim\exp(aL)$ in the gapped phase ($\abs{\Delta}>1$).
It would be worth sorting out in a broader setting
the connection between changes in the scaling behavior of $T$
and various types of phase transitions,
involving or not the appearance of many-body localized phases.

\ack

It is a pleasure to thank P Krapivsky, K Mallick and G Misguich for fruitful discussions,
and especially V Pasquier for his interest in this work
and for a careful reading of the manuscript.

\newpage

\section*{References}

\bibliography{paper.bib}

\end{document}